\documentclass[english,reprint,twocolumn,amsmath,amssymb,prl,nofootinbib,superscriptaddress]{revtex4-2}
\usepackage[T1]{fontenc}
\usepackage[latin9]{inputenc}
\usepackage{geometry}
\usepackage{graphicx}
\usepackage{bm}
\usepackage{hyperref}
\hypersetup{colorlinks, citecolor=Blue, linkcolor=Blue, urlcolor=Blue}
\usepackage[normalem]{ulem}
\usepackage[usenames,dvipsnames]{color}
\usepackage[sort&compress]{natbib}
\usepackage{textcomp}
\usepackage{babel}
\usepackage{notes2bib}

\bibnotesetup{
note-name = ,
use-sort-key = false
}

\AtBeginDocument{
   \setlength\abovedisplayskip{6pt}
   \setlength\belowdisplayskip{6pt}}

\begin{document}
\setlength{\parskip}{0pt}
\title{Milky Way Accelerometry via Millisecond Pulsar Timing}

\author{David~F.~Phillips}
\email{dphillips@cfa.harvard.edu}
\affiliation{Harvard-Smithsonian Center for Astrophysics, Cambridge, MA 02138, USA}

\author{Aakash~Ravi}
\email{aakash.ravi@gmail.com}
\affiliation{Quantum Technology Center, University of Maryland, College Park, MD 20742, USA}

\author{Reza~Ebadi}
\affiliation{Quantum Technology Center, University of Maryland, College Park, MD 20742, USA}
\affiliation{Department of Physics, University of Maryland, College Park, MD 20742, USA}

\author{Ronald~L.~Walsworth}
\affiliation{Quantum Technology Center, University of Maryland, College Park, MD 20742, USA}
\affiliation{Department of Physics, University of Maryland, College Park, MD 20742, USA}
\affiliation{Department of Electrical and Computer Engineering, University of Maryland, College Park, MD 20742, USA}

\date{\today}
\begin{abstract}
The temporal stability of millisecond pulsars is remarkable, rivaling even some terrestrial atomic clocks at long timescales. Using this property, we show that millisecond pulsars distributed in the galactic neighborhood form an ensemble of accelerometers from which we can directly extract the local galactic acceleration. From pulsar spin period measurements, we demonstrate acceleration sensitivity with about 1$\sigma$ precision using 117 pulsars. We also present a complementary analysis using orbital periods of 13 binary pulsar systems that eliminates systematics associated with pulsar braking and results in a local acceleration of $\left(1.7\pm0.5\right)\times{10}^{-10}\mathrm{ m/s}^2$ in good agreement with expectations. This work is a first step toward dynamically measuring acceleration gradients that will eventually inform us about the dark matter density distribution in the Milky Way galaxy.
\end{abstract}
\maketitle

\paragraph{Introduction ---}Stellar accelerations \cite{Ravi2019,Silverwood2019} were recently proposed as a promising alternative to kinematic observables in astrophysical characterizations of the dark matter distribution in the Milky Way galaxy. A key feature of such an acceleration-based approach is that it does not require the condition of dynamical equilibrium in the galaxy, an assumption that is inconsistent with evidence of disequilibria from astronomical surveys \cite{Newberg2002,Crane2003,Gomez2012,Widrow2012,Gomez2013,Carlin2013,Williams2013,Xu2015,Carrillo2018,Schoenrich2018,Antoja2018,Laporte2019,Bennett2019,Necib2019,Necib2019a}.\let\thefootnote\relax\footnote{$^{\ast,\dagger}$ These authors contributed equally to this work.}

Here, we present a complementary approach to Milky Way accelerometry based on long-term monitoring of the differential timing of electromagnetic emission from an ensemble of millisecond pulsars. The temporal stability of millisecond pulsars -- both due to their rotation (``spin'') and orbital motion (for binary pulsar systems) -- makes them very good astronomical clocks \cite{Hartnett2011}. This property has been useful for a variety of avenues of research \cite{Manchester2017}, including tests of general relativity \cite{Weisberg2010,Weisberg2016,Ransom2014,Archibald2018,Voisin2020}, gravitational wave detection \cite{Lommen2015,Shannon2013,Shannon2015,Demorest2013,Lentati2015,Babak2016,Demorest2013,Arzoumanian2016,Aggarwal2019,Verbiest2016}, characterizations of dark matter candidates such as massive gravitons \cite{Dubovsky2005,Baskaran2008,Pshirkov2008}, primordial black holes \cite{Seto2007,Kashiyama2012,Schutz2017,Dror2019}, and ultralight scalar fields \cite{Khmelnitsky2014,Porayko2014,Graham2016,DeMartino2017,DeMartino2018,Cai2018,Porayko2018,Caputo2019,Blas2020}. In addition, pulsar timing has been used to search for dark matter substructures \cite{Siegel2007,Baghram2011,Clark2015,Clark2015a,Ramani2020} in the galaxy such as minihalos.

\paragraph{Analysis framework ---} Here we present two analyses of the local acceleration as obtained from pulsar timing measurements. The first is based on pulsar spin periods and gives an acceleration of $\left(5.2\pm1.6\right)\times{10}^{-10}\mathrm{ m/s}^2$, a value which lies within $2\sigma$ of the expected acceleration. The second is derived from binary pulsar orbital periods, and gives an acceleration of $\left(1.7\pm0.5\right)\times{10}^{-10}\mathrm{ m/s}^2$. 

We begin by treating the pulsar observed spin period $P$, a quantity which depends on the pulsar's relative velocity with respect to the observer. In turn, the change of the observed spin period with time, $\dot{P}$, depends on the relative acceleration of the pulsar and observer, among other quantities. The apparent acceleration, $a$, can be related to  these quantities~\cite{Damour1991,Phinney1992,Nice1995} via 
\begin{equation}
a\equiv c\dot{P}/P-\mu^2 D=\Delta\vec{a}_{\mathrm{gal}}\cdot \hat{u}+ a_{\mathrm{br}},\label{eqn:accel_spin_p}
\end{equation}
where $c$ is the speed of light, $\mu$ is the total proper motion, and $D$ is the distance from the solar system barycenter (SSB) to the pulsar. In other words, $a$ is the apparent acceleration obtained from timing, $c\dot{P}/P$, corrected for perspective acceleration arising from the motion of the pulsar in the plane of the sky (known as the Shklovskii effect \cite{Shklovskii1970} in the pulsar community). Two components comprise $a$. First is the relative galactic acceleration of the pulsar with respect to the SSB, $\Delta\vec{a}_{\mathrm{gal}}$, projected onto the line of sight, $\hat{u}$, where $\hat{u}$ is a unit vector pointing from SSB to the pulsar. The second contribution to $a$, denoted $a_{\mathrm{br}}$, is the natural braking, or spin-down, of the pulsar rotation due to emission of electromagnetic radiation \cite{Lyne2012}.

Though the galactic term, $\Delta\vec{a}_{\mathrm{gal}}$, is the one of interest for Milky Way accelerometry, it cannot be separated from the braking contribution for an individual pulsar without additional information. However, with an ensemble of pulsars, we can statistically separate the two contributions if we treat the galactic contribution as a deterministic variable resulting from the galactic gravitational potential, and the braking acceleration as a random variable drawn from some probability distribution set by pulsar physics \cite{Xu2001}. Here, we assume that this braking distribution applies to all pulsars in our ensemble; and more generally, is the same throughout the galaxy.

In binary millisecond pulsar systems, a similar analysis to that developed for the spin period $P$ can be applied to the orbital period $P_{\mathrm{b}}$ \cite{Damour1991}. However, whereas the spin period decays due to the emission of electromagnetic radiation, the orbital period decays due to the emission of gravitational radiation. Fortunately, the change in orbital period due to gravitational waves can be determined directly if there is additional knowledge of the constituent masses $M_1,M_2$ and the orbital eccentricity $e$ \cite{Peters:1963}. Similar to equation \eqref{eqn:accel_spin_p}, the apparent acceleration can be written as
\begin{equation}
a_{\mathrm{b}}\equiv c\dot{P}_{\mathrm{b}}/P_{\mathrm{b}}-\mu^2 D - a_{\mathrm{GW}}=\Delta\vec{a}_{\mathrm{gal}}\cdot \hat{u},
\end{equation}
where $a_{\mathrm{GW}}$, the contribution from orbital period decay as the system emits gravitational waves, appears on the left hand side of the equation because it can be evaluated precisely for well-characterized pulsars, eliminating uncertainties associated with the spin period method  -- see Supplemental Material \cite{SM} for more details.

\begin{figure}
    \centering
    \includegraphics[width=\columnwidth]{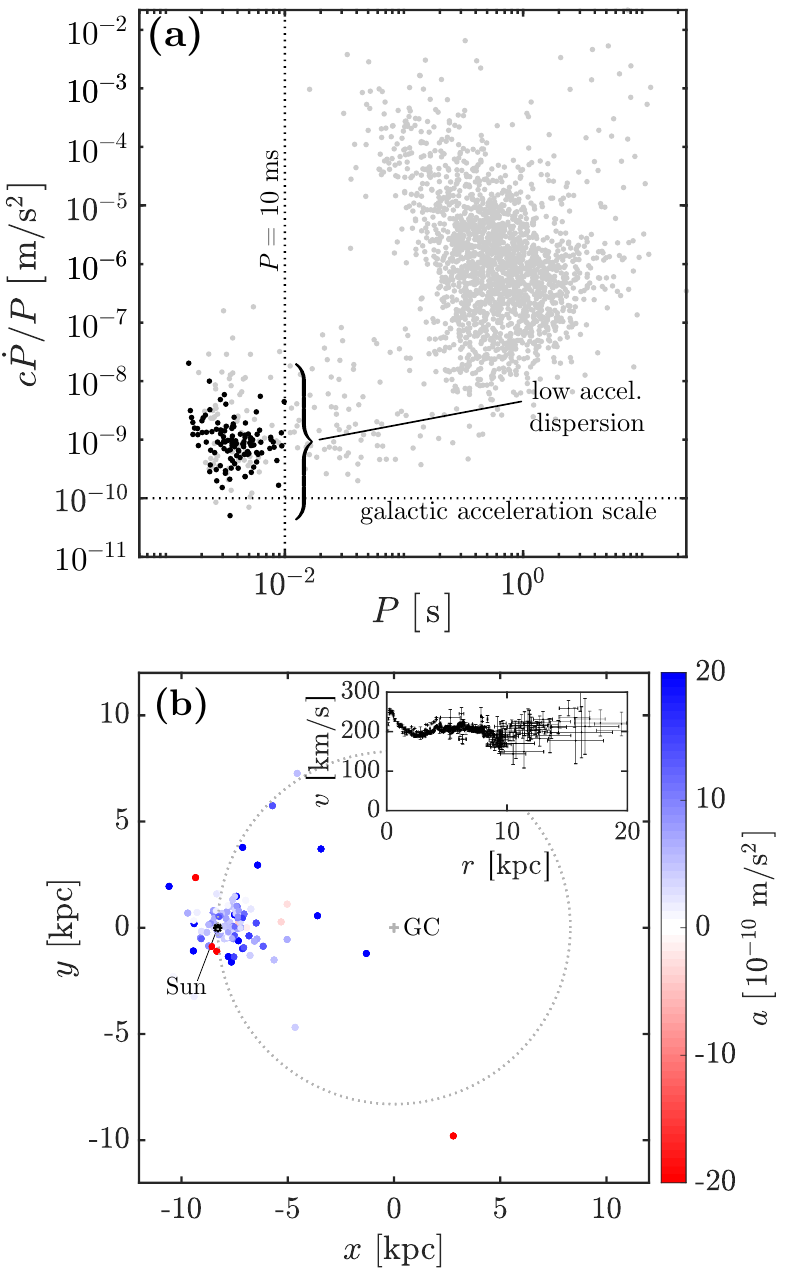}
    \caption{(a) Apparent pulsar acceleration ($c\dot{P}/P$) distribution vs. spin period $P$. We plot $c\dot{P}/P$ uncorrected for proper motion and visualize the entire pulsar catalogue. The pulsars selected for this work are in black in the bottom left of the plot. (b)~Spatial distribution of selected pulsars across the galactic midplane, color-coded by acceleration. The galactic center (GC) is marked with a cross ($+$). Pulsar distance uncertainties are suppressed for clarity. Inset: galactic rotation curve with data from Ref.~\cite{Sofue2009}.}
    \label{fig:selected_pulsars}
\end{figure}

We adopt a simple model \cite{Damour1991,Nice1995} for the galactic contribution to the apparent acceleration:
\begin{subequations}
\begin{align}
\Delta\vec{a}_{\mathrm{gal}}\cdot \hat{u} &\approx a_{0} f\left(D,l,b\right),\\
f\left(D,l,b\right) &\equiv -\cos b\left(\cos l+\frac{\beta}{\sin^{2}l+\beta^{2}}\right),
\end{align}
\label{eqn:dist}
\end{subequations}
where $a_{0}$ is the component of the SSB galactic acceleration toward the galactic center (GC) and $f$ is a function of the pulsar distance $D$, galactic longitude $l$ and latitude $b$, following the parametrization given in Ref. \cite{Damour1991}. The quantity $\beta$ is given by $\beta= \left(D/r_{0}\right)\cos b-\cos l$, where $r_0=8.1$ kpc is the SSB galactocentric radius. Here, we consider only planar galactic accelerations and neglect vertical accelerations, a choice we will justify below. We further assume a flat galactic rotation curve. As shown in the inset of Fig.~\ref{fig:selected_pulsars}b, the galactic rotation curve, inferred from diverse astronomical observations \cite{Sofue2009}, deviates from flatness only within a few kpc of the GC; very few pulsars have been characterized in this region, enabling us to use the flat galactic rotation curve approximation.

For this model and assumptions, the apparent acceleration $a$ is linear in the pulsar position function $f$, with slope giving the local galactic acceleration $a_{0}$. Note that this framework could be straightforwardly extended to study deviations from linearity, which could inform us about the local dark matter density and its distribution in the galaxy \cite{Read2014}. 

\paragraph{Spin period analysis---} Our pulsar timing data is sourced from the ATNF pulsar catalogue \cite{Manchester2005}. We select pulsars satisfying the following criteria:
\begin{enumerate}
\setlength\itemsep{-0.5em}
    \item Period $P<10$  ms.
    \item Measured total proper motion, $\mu$.
    \item Pulsar is not in a globular cluster.
\end{enumerate}
Criterion 1 provides a conservative basis for selecting millisecond pulsars; we choose this subset of all pulsars due to their lower acceleration dispersion as seen in panel (a) of Fig.~\ref{fig:selected_pulsars} and their greater age compared to more slowly-rotating pulsars; below, we describe the results of relaxing this constraint. Criterion 2 enables proper motion corrections to pulsar accelerations; these can be as large or larger than the galactic contribution of interest here. Criterion 3 ensures that globular cluster dynamics are not confused for galactic dynamics. The selected pulsar sample is shown in Figure \ref{fig:selected_pulsars}. A total of 117 pulsars satisfy all our criteria. The inset of panel~(b), which shows the galactic rotation curve data from Ref.~\cite{Sofue2009}, justifies our flat rotation curve assumption.

\begin{figure}
    \centering
    \includegraphics[width=\columnwidth]{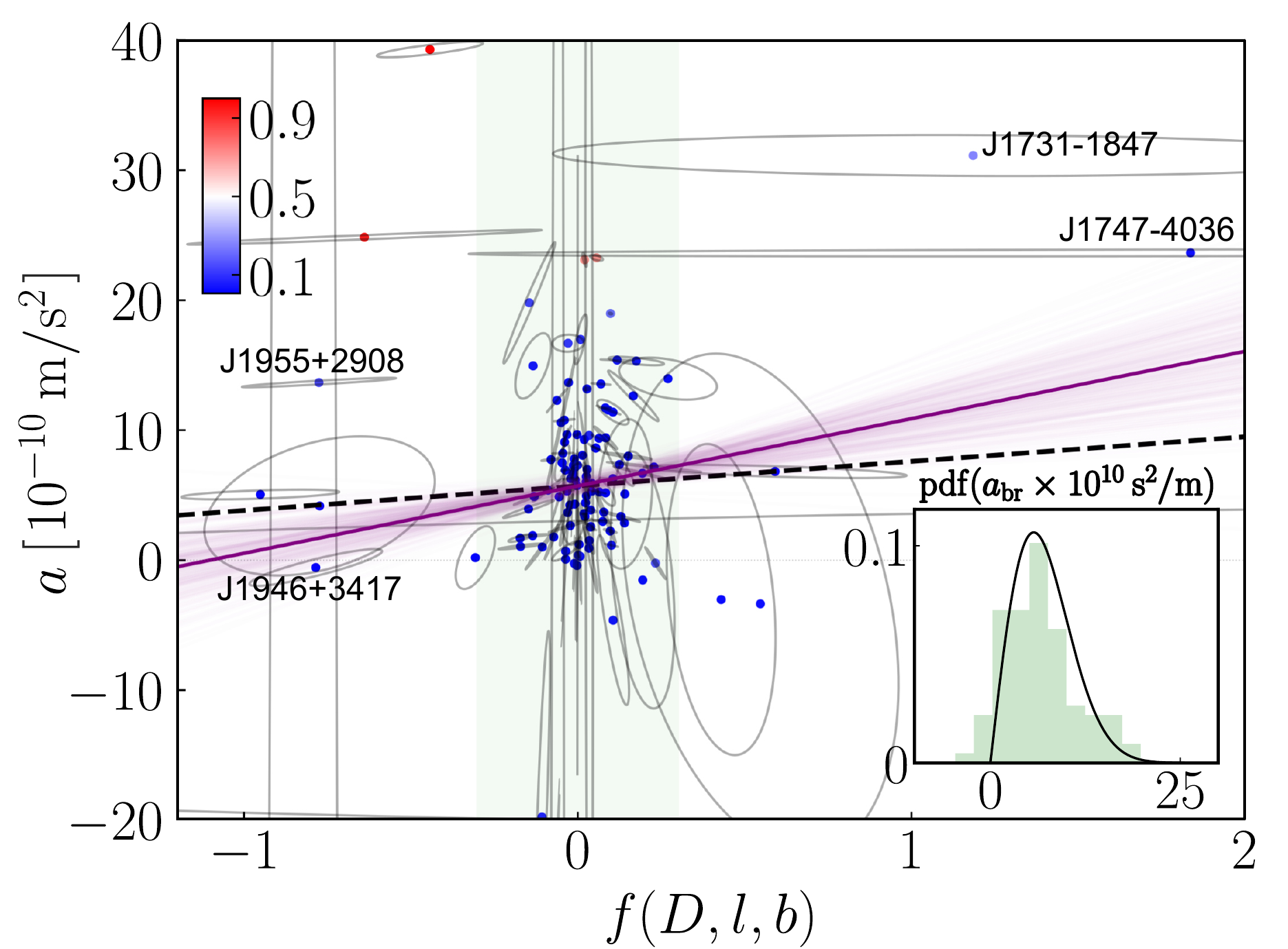}
    \caption{Local galactic acceleration extracted from pulsar spin periods for selected pulsars. The abscissa $f\left(D,l,b\right)$ is a function of pulsar distance $D$, galactic longitude $l$ and latitude $b$ (Eq.~\ref{eqn:dist}). Each dot represents an individual pulsar and the ellipse around the dot represents the observational covariance matrix for that pulsar. The points are color-coded by the degree of certainty that a particular point is an outlier \cite{SM}. The thick violet line is the best-fit estimate given by the relationship $a=\hat{a}_0f+\hat{\lambda}$, where $\hat{a}_0=\left(5.2\pm1.6\right)\times{10}^{-10}\mathrm{ m/s}^2$ and $\hat{\lambda}=\left(5.7\pm0.4\right)\times{10}^{-10}\mathrm{ m/s}^2$. The fainter violet lines around the thick violet represent the distribution of fitted accelerations \cite{SM}. Lastly, the dashed black line has slope $2.2 \times{10}^{-10}\text{ m/s}^2$, which is the nominal value for the solar system barycenter assuming a circular velocity of 233~km/s \cite{Evans:2018bqy} and galactocentric distance of 8.1 kpc \cite{Evans:2018bqy}. (Inset) Recovered braking distribution -- in black, the distribution described by $\hat{\lambda}$ is plotted over a histogram of $a$ values for pulsars within the green region $\left|f\right|<0.3$ in (a) that have an outlier score $<0.5$. In this region, the braking effect dominates over the galactic component.}
    \label{fig:results}
\end{figure}

Using information from the pulsar catalogue, we prepare $a$ vs.\ $f$ data for fitting as described in the Supplemental Material~\cite{SM}. We employ a Bayesian mixture model \cite{SM} to extract the local galactic acceleration $a_0$ as well as the braking distribution. The mixture model is capable of handling statistical outliers; and in practice, it rejects approximately 10\% of the pulsars in our sample. See Table I of \cite{SM} for outlier scores. The braking distribution is modelled as a Rayleigh distribution with a single parameter $\lambda$ because of its simplicity and distinct positive-skew (a feature observed in the data). Furthermore, it has the property of being supported on only nonnegative real numbers, therefore encoding our physical notion that pulsars spin down. The result of our analysis is shown in Figure~\ref{fig:results}.

The best-fit estimate for the local galactic acceleration $\hat{a}_0=\left(5.2\pm1.6\right)\times{10}^{-10}\mathrm{ m/s}^2$ is in tension at the $\sim 2\sigma$ level with $(2.2 \pm 0.1)\times{10}^{-10}\text{ m/s}^2$, the value we would expect from a circular velocity of $\left(233 \pm 3\right)$~km/s \cite{McMillan2016} at the SSB galactocentric distance of 8.2 kpc \cite{abuter_geometric_2019}. This modest disagreement is at least partly driven by limitations of the existing dataset. The first limitation is that the distribution of pulsars in $f$ is primarily clustered about $f=0$ with only a few at large $\left|f\right|$. The second limitation is that the slope is sensitive to the details of outlier removal. To quantify this sensitivity, we performed a dropout study where we repeated the fit procedure many times, removing a different pulsar from the dataset each time. This analysis revealed that only a handful of pulsars (the four points marked with pulsar alphanumeric identifiers in Figure \ref{fig:results}) have sufficient leverage to change the slope by more than 10\%. Of these four pulsars, J1731--1847 has the largest effect. This pulsar is an eclipsing binary system that is susceptible to changes in orbital period as well as dispersion measure \cite{Bates2011}, thereby reducing the quality of both its timing data and distance estimate. Removing J1731--1847 as well as the lone pulsar J1747--4036 (which has an extremely large distance uncertainty) results in $\hat{a}_0\rightarrow\left(2.5\pm2.0\right)\times{10}^{-10}\mathrm{ m/s}^2$, consistent with the expected value from the Milky Way rotation curve.  A complete characterization of the properties of all 117 pulsars used in this analysis is beyond the scope of our current work. 

The inset in Figure \ref{fig:results}, demonstrates our ability to simultaneously fit the pulsar braking distribution and the galactic acceleration. The black curve is the Rayleigh distribution given by $\hat{\lambda}$, which agrees well with the histogram of pulsar apparent accelerations within the SSB vicinity (i.e., $\left|f\right|<0.3$) that have an outlier score $<0.5$. Since the pulsars in the vicinity are co-accelerating with the SSB, the galactic contribution is small compared to the braking contribution, and therefore, these pulsars serve as a test of our model. We stress that the black curve is not a fit to the histogram but is derived from the entire pulsar ensemble and is superimposed on the histogram for comparison. One caveat of note is that the tail of the true braking distribution may be longer than our model suggests. This is because the outlier pruning in the Bayesian mixture model removes pulsars with very large braking values. Thus, the scope of our distribution is limited to only high-$Q$ (quality factor) millisecond pulsars ($a_{\mathrm{br}}\lesssim20\times{10}^{-10}\mathrm{ m/s}^2$) and does not represent lossy pulsars.

To estimate the effect of vertical accelerations, we use a disk+halo model \cite{Kuijken1989}, which gives the vertical component of the galactic acceleration as \mbox{$a_z=-3.27\left(1.25\tilde{z}/\sqrt{\tilde{z}^2+0.0324}+0.58\tilde{z}\right)\times{10}^{-11}\mathrm{ m/s}^2$}. The symbol $\tilde{z}$ is a dimensionless vertical distance given by $z/z_s$, and $z_s=1$ kpc. The vertical contribution to $\Delta\vec{a}_{\mathrm{gal}}\cdot \hat{u}$ is $a_z\sin b$; and in most cases, this turns out to be only a few percent of the planar component. We re-ran the analysis with this component removed and found no appreciable change in the fitted parameters or parameter uncertainties, justifying our neglect of the vertical accelerations.

We also relaxed Criterion 1 to allow all pulsars with $P<100$~ms and $c\dot{P}/P<10^{-6}$~m/s${}^2$ (see Fig.~\ref{fig:selected_pulsars}a) into our sample. This increases our usable number of pulsars by $\sim$25\%, but has little effect on our parameter estimates or uncertainties.

Looking forward to future surveys of millisecond pulsars, we highlight pulsars within our sample that would benefit from improved measurements in Table I of \cite{SM}. We also explore how the acceleration sensitivity $\sigma_{a_0}$ scales with the number of pulsars $N$ in the sample. A Monte Carlo simulation \cite{SM} reveals that with 100 pulsars, our spin period analysis provides sensitivity to the local galactic acceleration at a $\sim$1$\sigma$ level (consistent with our fit uncertainties); and we can expect to reach $\sim$3$\sigma$ level sensitivity when we have approximately 1,000 pulsars. These extrapolations to larger numbers of pulsars are limited by the biases implicit in the distribution of currently well-characterized pulsars, as well as the unknown distribution of pulsar locations and braking parameters to be observed in the future.

\begin{figure}
    \centering
    \includegraphics[width=\columnwidth]{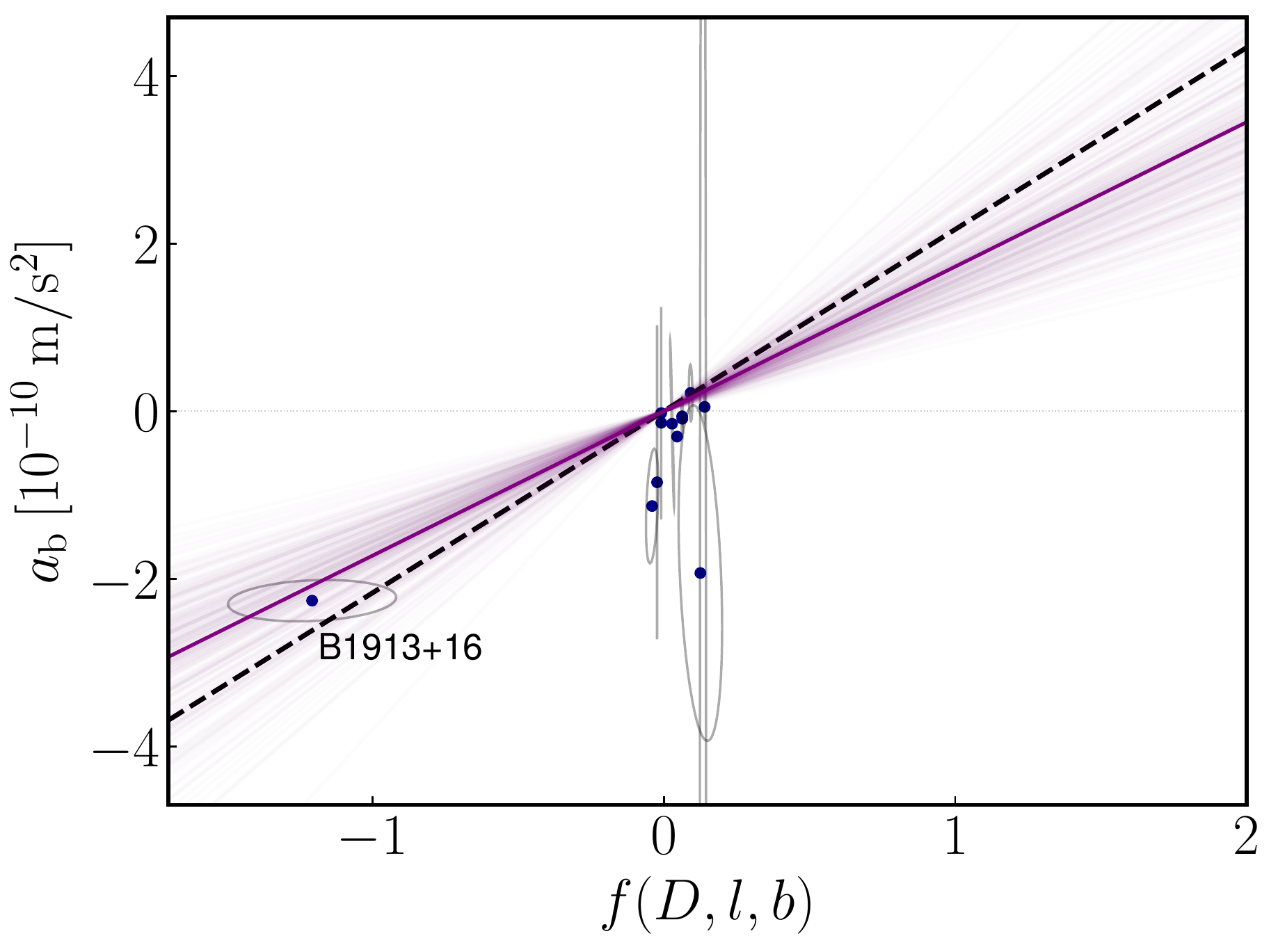}
    \caption {Local galactic acceleration extracted from binary millisecond pulsar orbital periods given in Table II of Ref. \cite{SM}. The abscissa $f\left(D,l,b\right)$ is a function of pulsar distance $D$, galactic longitude $l$ and latitude $b$. Each dot represents an individual binary pulsar system and the ellipse around the dot represents the observational covariance matrix for that system. The thick violet line is the best-fit estimate given by the relationship $a_{\mathrm{b}}=\hat{a}_0f$, where $\hat{a}_0=\left(1.73^{+0.50}_{-0.37}\right)\times{10}^{-10}\mathrm{ m/s}^2$. The fainter violet lines around the thick violet represent the distribution of fitted accelerations \cite{SM}. Finally, the dashed black line has slope $2.2 \times{10}^{-10}\text{ m/s}^2$, which is the nominal value for the solar system barycenter assuming a circular velocity of 233~km/s \cite{Evans:2018bqy} and galactocentric distance of 8.1~kpc \cite{Evans:2018bqy}.}
    \label{fig:results_orbital}
\end{figure}

\paragraph{Orbital period analysis ---} For our second analysis method, binary millisecond pulsar orbital periods serve as our reference clocks. A summary of the relevant data for the 13 binary pulsars used in our analysis is presented in Table II of the Supplemental Material \cite{SM}. In the spin period analysis above, the distribution of intrinsic spin-down rates must be inferred \emph{statistically}. In contrast, for binary pulsars with well-characterized orbits, we can \emph{directly} account for the change of the orbital period of the pulsar system due to gravitational wave emission ~\cite{Peters:1963, Shao:2020fka, SM}. Once such corrections are applied, the fitting procedure \cite{SM} is very similar to the spin periods. The result, as shown in Figure \ref{fig:results_orbital}, gives a best-fit estimate for the local galactic acceleration of $\hat{a}_0=\left(1.73^{+0.49}_{-0.37}\right)\times{10}^{-10}\mathrm{ m/s}^2$. This result is consistent with the expected value from the circular motion assumption, i.e., $\left(2.2 \pm 0.1\right)\times{10}^{-10}\text{ m/s}^2$. Note that the result critically depends on the Hulse-Taylor binary pulsar (B1913+16); removing it from the analysis leads to a result for the local acceleration of $\left(-0.55 \pm 1.3\right)\times{10}^{-10}\text{ m/s}^2$, which has threefold larger uncertainty than the result including this pulsar, and differs by about $2\sigma$ from the nominally expected value. Additional high-precision measurements of the constituent masses for distant binary pulsar systems will be needed to verify the significance of the present result. Nonetheless, the orbital period analysis is a promising approach, as the explicit compensation for gravitational wave energy loss enables reaching a given acceleration sensitivity with far fewer pulsars than for the spin period analysis. As with spin periods, we highlight pulsars that would benefit from improved measurements in Table II of \cite{SM}.


\paragraph{Conclusions and outlook ---} In summary, we demonstrated two techniques for extracting the acceleration of the solar system barycenter in the Milky Way gravitational potential using pulsar timing measurements. These methods are dynamical: they directly access the relative acceleration of pulsars in the galactic neighbourhood. Such techniques are complementary to existing kinematic approaches, which currently attain higher precision for the Milky Way gravitational potential, but do so with the assumption of dynamical equilibrium. Using spin periods for 117 millisecond pulsars, we reach a sensitivity to the local galactic acceleration with $\sim$1$\sigma$ precision, but are limited by both the spatial distribution of well-characterized pulsars and sensitivity to outliers. When high-quality data becomes available for $\sim$1,000 millisecond pulsars, we expect to reach $\sim$3$\sigma$ precision with this approach. Additionally, we show local galactic acceleration sensitivity in agreement ($1\sigma$ precision) with the nominal value from the galactic rotation curve by analyzing the orbital periods of just 13 binary pulsars. Orbital periods, being unaffected by pulsar braking, offer a cleaner approach to determining the local galactic acceleration. 

In the local neighborhood, dark matter comprises approximately 13\% of the total density \cite{Mckee2015}. Thus, under the assumption of a flattened galactic potential, a 3-4\% fractional uncertainty of the local acceleration coupled with improved accuracy in the baryon density (which is only currently known to $\sim$15\%) are required to achieve 3$\sigma$ precision on the local dark matter density.

With both the spin and orbital period approaches, additional and improved millisecond pulsar data will provide better galactic acceleration sensitivity, thereby enabling searches for halo substructure~\cite{Buckley:2018,Siegel2007} on the several kpc scale along with deviations from the flat galactic rotation curve assumed in our current analysis. For example, future releases of \emph{Gaia} data, including proper motion measurements of pulsar systems and improved distance measurements, could be extremely valuable in extending the reach of the current dataset (see Ref. \cite{Jennings2018,Mingarelli2018} for examples).  Additional binary millisecond pulsars characterized by pulsar timing arrays~\cite{Lommen2015,Shannon2013} will also improve the sensitivity of the orbital period technique. We hope our efforts to determine the galactic gravitational potential dynamically through pulsar timing, as well as characterize the statistical properties of pulsar spin-down rates, will provide motivations for high-precision pulsar astronomy in the future.



\vspace{1cm}\raggedbottom

\begin{acknowledgments}
We thank N. Langellier and T. W. Milbourne for valuable technical advice on Bayesian regression and MCMC sampling. We are also grateful to M.~J.~Turner for insightful discussions, and S. Rajendran for a careful read of the manuscript. This work was supported by the DOE QuANTISED program under Award No.\ DE-SC0019396; the Army Research Laboratory MAQP program under Contract No.\ W911NF-19-2-0181; and the University of Maryland Quantum Technology Center. Our analysis codes and data are available online \bibnote{\href{ https://github.com/R3zaEbadi/MW-Accelerometry}{\tt https://github.com/R3zaEbadi/MW-Accelerometry}}. In our code, we make use of the following Python packages: {\tt psrqpy} \cite{psrqpy}, {\tt emcee} \cite{DFM2013}, {\tt corner} \cite{corner}.

After submission of our manuscript, a related paper was posted to the arXiv and subsequently published \cite{Chakrabarti2021}.  This paper reports a similar analysis of binary pulsar orbital periods to the present Letter, with results of comparable precision. The authors also extend the analysis to include vertical accelerations, and make estimates the galactic midplane mass density, the local dark matter density and the slope of the vertical acceleration profile. Ref. \cite{Chakrabarti2021} cites our arXiv preprint, but incorrectly describes our paper as being limited to spin periods and also mischaracterizes our analysis of the systematic errors associated with the spin period method.

Also after submission of our manuscript, the Sun's acceleration in the galaxy was determined to be $\left(2.32\pm0.16\right)\times10^{-10}$~m/s$^{2}$ \cite{Klioner2020} using precision \emph{Gaia} astrometry data for $>10^{6}$ compact extragalactic sources. While this measurement is more precise than those reported here, we envision that additional pulsar data will enable us to both improve our precision and probe galactic structure beyond the local acceleration. It is also worth noting that the astrometry-based measurement of the local acceleration has already been combined with binary pulsar orbital period measurements to extract galactic parameters \cite{Bovy2020}, highlighting the complementarity of these approaches.

\end{acknowledgments}

\bibliographystyle{apsrev4-1}
\bibliography{references}

\begin{thebibliography}{97}%
\makeatletter
\providecommand \@ifxundefined [1]{%
 \@ifx{#1\undefined}
}%
\providecommand \@ifnum [1]{%
 \ifnum #1\expandafter \@firstoftwo
 \else \expandafter \@secondoftwo
 \fi
}%
\providecommand \@ifx [1]{%
 \ifx #1\expandafter \@firstoftwo
 \else \expandafter \@secondoftwo
 \fi
}%
\providecommand \natexlab [1]{#1}%
\providecommand \enquote  [1]{``#1''}%
\providecommand \bibnamefont  [1]{#1}%
\providecommand \bibfnamefont [1]{#1}%
\providecommand \citenamefont [1]{#1}%
\providecommand \href@noop [0]{\@secondoftwo}%
\providecommand \href [0]{\begingroup \@sanitize@url \@href}%
\providecommand \@href[1]{\@@startlink{#1}\@@href}%
\providecommand \@@href[1]{\endgroup#1\@@endlink}%
\providecommand \@sanitize@url [0]{\catcode `\\12\catcode `\$12\catcode
  `\&12\catcode `\#12\catcode `\^12\catcode `\_12\catcode `\%12\relax}%
\providecommand \@@startlink[1]{}%
\providecommand \@@endlink[0]{}%
\providecommand \url  [0]{\begingroup\@sanitize@url \@url }%
\providecommand \@url [1]{\endgroup\@href {#1}{\urlprefix }}%
\providecommand \urlprefix  [0]{URL }%
\providecommand \Eprint [0]{\href }%
\providecommand \doibase [0]{http://dx.doi.org/}%
\providecommand \selectlanguage [0]{\@gobble}%
\providecommand \bibinfo  [0]{\@secondoftwo}%
\providecommand \bibfield  [0]{\@secondoftwo}%
\providecommand \translation [1]{[#1]}%
\providecommand \BibitemOpen [0]{}%
\providecommand \bibitemStop [0]{}%
\providecommand \bibitemNoStop [0]{.\EOS\space}%
\providecommand \EOS [0]{\spacefactor3000\relax}%
\providecommand \BibitemShut  [1]{\csname bibitem#1\endcsname}%
\let\auto@bib@innerbib\@empty
\bibitem [{\citenamefont {Ravi}\ \emph {et~al.}(2019)\citenamefont {Ravi},
  \citenamefont {Langellier}, \citenamefont {Phillips}, \citenamefont
  {Buschmann}, \citenamefont {Safdi},\ and\ \citenamefont
  {Walsworth}}]{Ravi2019}%
  \BibitemOpen
  \bibfield  {author} {\bibinfo {author} {\bibfnamefont {A.}~\bibnamefont
  {Ravi}}, \bibinfo {author} {\bibfnamefont {N.}~\bibnamefont {Langellier}},
  \bibinfo {author} {\bibfnamefont {D.~F.}\ \bibnamefont {Phillips}}, \bibinfo
  {author} {\bibfnamefont {M.}~\bibnamefont {Buschmann}}, \bibinfo {author}
  {\bibfnamefont {B.~R.}\ \bibnamefont {Safdi}}, \ and\ \bibinfo {author}
  {\bibfnamefont {R.~L.}\ \bibnamefont {Walsworth}},\ }\href {\doibase gf9t28}
  {\bibfield  {journal} {\bibinfo  {journal} {Phys. Rev. Lett.}\ }\textbf
  {\bibinfo {volume} {123}},\ \bibinfo {pages} {091101} (\bibinfo {year}
  {2019})}\BibitemShut {NoStop}%
\bibitem [{\citenamefont {Silverwood}\ and\ \citenamefont
  {Easther}(2019)}]{Silverwood2019}%
  \BibitemOpen
  \bibfield  {author} {\bibinfo {author} {\bibfnamefont {H.}~\bibnamefont
  {Silverwood}}\ and\ \bibinfo {author} {\bibfnamefont {R.}~\bibnamefont
  {Easther}},\ }\href {\doibase 10.1017/pasa.2019.25} {\bibfield  {journal}
  {\bibinfo  {journal} {Publ. Astron. Soc. Aust.}\ }\textbf {\bibinfo {volume}
  {36}},\ \bibinfo {pages} {e038} (\bibinfo {year} {2019})}\BibitemShut
  {NoStop}%
\bibitem [{\citenamefont {Newberg}\ \emph {et~al.}(2002)\citenamefont
  {Newberg}, \citenamefont {Yanny}, \citenamefont {Rockosi}, \citenamefont
  {Grebel}, \citenamefont {Rix}, \citenamefont {Brinkmann}, \citenamefont
  {Csabai}, \citenamefont {Hennessy}, \citenamefont {Hindsley}, \citenamefont
  {Ibata} \emph {et~al.}}]{Newberg2002}%
  \BibitemOpen
  \bibfield  {author} {\bibinfo {author} {\bibfnamefont {H.~J.}\ \bibnamefont
  {Newberg}}, \bibinfo {author} {\bibfnamefont {B.}~\bibnamefont {Yanny}},
  \bibinfo {author} {\bibfnamefont {C.}~\bibnamefont {Rockosi}}, \bibinfo
  {author} {\bibfnamefont {E.~K.}\ \bibnamefont {Grebel}}, \bibinfo {author}
  {\bibfnamefont {H.-W.}\ \bibnamefont {Rix}}, \bibinfo {author} {\bibfnamefont
  {J.}~\bibnamefont {Brinkmann}}, \bibinfo {author} {\bibfnamefont
  {I.}~\bibnamefont {Csabai}}, \bibinfo {author} {\bibfnamefont
  {G.}~\bibnamefont {Hennessy}}, \bibinfo {author} {\bibfnamefont {R.~B.}\
  \bibnamefont {Hindsley}}, \bibinfo {author} {\bibfnamefont {R.}~\bibnamefont
  {Ibata}},  \emph {et~al.},\ }\href {\doibase 10.1086/338983} {\bibfield
  {journal} {\bibinfo  {journal} {Astrophys. J.}\ }\textbf {\bibinfo {volume}
  {569}},\ \bibinfo {pages} {245} (\bibinfo {year} {2002})}\BibitemShut
  {NoStop}%
\bibitem [{\citenamefont {Crane}\ \emph {et~al.}(2003)\citenamefont {Crane},
  \citenamefont {Majewski}, \citenamefont {Rocha-Pinto}, \citenamefont
  {Frinchaboy}, \citenamefont {Skrutskie},\ and\ \citenamefont
  {Law}}]{Crane2003}%
  \BibitemOpen
  \bibfield  {author} {\bibinfo {author} {\bibfnamefont {J.~D.}\ \bibnamefont
  {Crane}}, \bibinfo {author} {\bibfnamefont {S.~R.}\ \bibnamefont {Majewski}},
  \bibinfo {author} {\bibfnamefont {H.~J.}\ \bibnamefont {Rocha-Pinto}},
  \bibinfo {author} {\bibfnamefont {P.~M.}\ \bibnamefont {Frinchaboy}},
  \bibinfo {author} {\bibfnamefont {M.~F.}\ \bibnamefont {Skrutskie}}, \ and\
  \bibinfo {author} {\bibfnamefont {D.~R.}\ \bibnamefont {Law}},\ }\href
  {\doibase 10.1086/378767} {\bibfield  {journal} {\bibinfo  {journal}
  {Astrophys. J. Lett.}\ }\textbf {\bibinfo {volume} {594}},\ \bibinfo {pages}
  {L119} (\bibinfo {year} {2003})}\BibitemShut {NoStop}%
\bibitem [{\citenamefont {G\'{o}mez}\ \emph {et~al.}(2012)\citenamefont
  {G\'{o}mez}, \citenamefont {Minchev}, \citenamefont {O'Shea}, \citenamefont
  {Lee}, \citenamefont {Beers}, \citenamefont {An}, \citenamefont {Bullock},
  \citenamefont {Purcell},\ and\ \citenamefont {Villalobos}}]{Gomez2012}%
  \BibitemOpen
  \bibfield  {author} {\bibinfo {author} {\bibfnamefont {F.~A.}\ \bibnamefont
  {G\'{o}mez}}, \bibinfo {author} {\bibfnamefont {I.}~\bibnamefont {Minchev}},
  \bibinfo {author} {\bibfnamefont {B.~W.}\ \bibnamefont {O'Shea}}, \bibinfo
  {author} {\bibfnamefont {Y.~S.}\ \bibnamefont {Lee}}, \bibinfo {author}
  {\bibfnamefont {T.~C.}\ \bibnamefont {Beers}}, \bibinfo {author}
  {\bibfnamefont {D.}~\bibnamefont {An}}, \bibinfo {author} {\bibfnamefont
  {J.~S.}\ \bibnamefont {Bullock}}, \bibinfo {author} {\bibfnamefont {C.~W.}\
  \bibnamefont {Purcell}}, \ and\ \bibinfo {author} {\bibfnamefont
  {A.}~\bibnamefont {Villalobos}},\ }\href {\doibase f34sst} {\bibfield
  {journal} {\bibinfo  {journal} {Mon. Not. R. Astron. Soc.}\ }\textbf
  {\bibinfo {volume} {423}},\ \bibinfo {pages} {3727} (\bibinfo {year}
  {2012})}\BibitemShut {NoStop}%
\bibitem [{\citenamefont {Widrow}\ \emph {et~al.}(2012)\citenamefont {Widrow},
  \citenamefont {Gardner}, \citenamefont {Yanny}, \citenamefont {Dodelson},\
  and\ \citenamefont {Chen}}]{Widrow2012}%
  \BibitemOpen
  \bibfield  {author} {\bibinfo {author} {\bibfnamefont {L.~M.}\ \bibnamefont
  {Widrow}}, \bibinfo {author} {\bibfnamefont {S.}~\bibnamefont {Gardner}},
  \bibinfo {author} {\bibfnamefont {B.}~\bibnamefont {Yanny}}, \bibinfo
  {author} {\bibfnamefont {S.}~\bibnamefont {Dodelson}}, \ and\ \bibinfo
  {author} {\bibfnamefont {H.-Y.}\ \bibnamefont {Chen}},\ }\href
  {https://iopscience.iop.org/article/10.1088/2041-8205/750/2/L41} {\bibfield
  {journal} {\bibinfo  {journal} {Astrophys. J. Lett.}\ }\textbf {\bibinfo
  {volume} {750}},\ \bibinfo {pages} {L41} (\bibinfo {year}
  {2012})}\BibitemShut {NoStop}%
\bibitem [{\citenamefont {G\'{o}mez}\ \emph {et~al.}(2013)\citenamefont
  {G\'{o}mez}, \citenamefont {Minchev}, \citenamefont {O'Shea}, \citenamefont
  {Beers}, \citenamefont {Bullock},\ and\ \citenamefont {Purcell}}]{Gomez2013}%
  \BibitemOpen
  \bibfield  {author} {\bibinfo {author} {\bibfnamefont {F.~A.}\ \bibnamefont
  {G\'{o}mez}}, \bibinfo {author} {\bibfnamefont {I.}~\bibnamefont {Minchev}},
  \bibinfo {author} {\bibfnamefont {B.~W.}\ \bibnamefont {O'Shea}}, \bibinfo
  {author} {\bibfnamefont {T.~C.}\ \bibnamefont {Beers}}, \bibinfo {author}
  {\bibfnamefont {J.~S.}\ \bibnamefont {Bullock}}, \ and\ \bibinfo {author}
  {\bibfnamefont {C.~W.}\ \bibnamefont {Purcell}},\ }\href {\doibase f4xd3s}
  {\bibfield  {journal} {\bibinfo  {journal} {Mon. Not. R. Astron. Soc.}\
  }\textbf {\bibinfo {volume} {429}},\ \bibinfo {pages} {159} (\bibinfo {year}
  {2013})}\BibitemShut {NoStop}%
\bibitem [{\citenamefont {Carlin}\ \emph {et~al.}(2013)\citenamefont {Carlin},
  \citenamefont {DeLaunay}, \citenamefont {Newberg}, \citenamefont {Deng},
  \citenamefont {Gole}, \citenamefont {Grabowski}, \citenamefont {Jin},
  \citenamefont {Liu}, \citenamefont {Liu}, \citenamefont {Luo} \emph
  {et~al.}}]{Carlin2013}%
  \BibitemOpen
  \bibfield  {author} {\bibinfo {author} {\bibfnamefont {J.~L.}\ \bibnamefont
  {Carlin}}, \bibinfo {author} {\bibfnamefont {J.}~\bibnamefont {DeLaunay}},
  \bibinfo {author} {\bibfnamefont {H.~J.}\ \bibnamefont {Newberg}}, \bibinfo
  {author} {\bibfnamefont {L.}~\bibnamefont {Deng}}, \bibinfo {author}
  {\bibfnamefont {D.}~\bibnamefont {Gole}}, \bibinfo {author} {\bibfnamefont
  {K.}~\bibnamefont {Grabowski}}, \bibinfo {author} {\bibfnamefont
  {G.}~\bibnamefont {Jin}}, \bibinfo {author} {\bibfnamefont {C.}~\bibnamefont
  {Liu}}, \bibinfo {author} {\bibfnamefont {X.}~\bibnamefont {Liu}}, \bibinfo
  {author} {\bibfnamefont {A.-L.}\ \bibnamefont {Luo}},  \emph {et~al.},\
  }\href {\doibase 10.1088/2041-8205/777/1/L5} {\bibfield  {journal} {\bibinfo
  {journal} {Astrophys. J. Lett.}\ }\textbf {\bibinfo {volume} {777}},\
  \bibinfo {pages} {L5} (\bibinfo {year} {2013})}\BibitemShut {NoStop}%
\bibitem [{\citenamefont {Williams}\ \emph {et~al.}(2013)\citenamefont
  {Williams}, \citenamefont {Steinmetz}, \citenamefont {Binney}, \citenamefont
  {Siebert}, \citenamefont {Enke}, \citenamefont {Famaey}, \citenamefont
  {Minchev}, \citenamefont {de~Jong}, \citenamefont {Boeche}, \citenamefont
  {Freeman} \emph {et~al.}}]{Williams2013}%
  \BibitemOpen
  \bibfield  {author} {\bibinfo {author} {\bibfnamefont {M.~E.~K.}\
  \bibnamefont {Williams}}, \bibinfo {author} {\bibfnamefont {M.}~\bibnamefont
  {Steinmetz}}, \bibinfo {author} {\bibfnamefont {J.}~\bibnamefont {Binney}},
  \bibinfo {author} {\bibfnamefont {A.}~\bibnamefont {Siebert}}, \bibinfo
  {author} {\bibfnamefont {H.}~\bibnamefont {Enke}}, \bibinfo {author}
  {\bibfnamefont {B.}~\bibnamefont {Famaey}}, \bibinfo {author} {\bibfnamefont
  {I.}~\bibnamefont {Minchev}}, \bibinfo {author} {\bibfnamefont {R.~S.}\
  \bibnamefont {de~Jong}}, \bibinfo {author} {\bibfnamefont {C.}~\bibnamefont
  {Boeche}}, \bibinfo {author} {\bibfnamefont {K.~C.}\ \bibnamefont {Freeman}},
   \emph {et~al.},\ }\href {\doibase 10.1093/mnras/stt1522} {\bibfield
  {journal} {\bibinfo  {journal} {Mon. Not. R. Astron. Soc.}\ }\textbf
  {\bibinfo {volume} {436}},\ \bibinfo {pages} {101} (\bibinfo {year}
  {2013})}\BibitemShut {NoStop}%
\bibitem [{\citenamefont {Xu}\ \emph {et~al.}(2015)\citenamefont {Xu},
  \citenamefont {Newberg}, \citenamefont {Carlin}, \citenamefont {Liu},
  \citenamefont {Deng}, \citenamefont {Li}, \citenamefont {Sch\"{o}nrich},\
  and\ \citenamefont {Yanny}}]{Xu2015}%
  \BibitemOpen
  \bibfield  {author} {\bibinfo {author} {\bibfnamefont {Y.}~\bibnamefont
  {Xu}}, \bibinfo {author} {\bibfnamefont {H.~J.}\ \bibnamefont {Newberg}},
  \bibinfo {author} {\bibfnamefont {J.~L.}\ \bibnamefont {Carlin}}, \bibinfo
  {author} {\bibfnamefont {C.}~\bibnamefont {Liu}}, \bibinfo {author}
  {\bibfnamefont {L.}~\bibnamefont {Deng}}, \bibinfo {author} {\bibfnamefont
  {J.}~\bibnamefont {Li}}, \bibinfo {author} {\bibfnamefont {R.}~\bibnamefont
  {Sch\"{o}nrich}}, \ and\ \bibinfo {author} {\bibfnamefont {B.}~\bibnamefont
  {Yanny}},\ }\href {\doibase 10.1088/0004-637X/801/2/105} {\bibfield
  {journal} {\bibinfo  {journal} {Astrophys. J.}\ }\textbf {\bibinfo {volume}
  {801}},\ \bibinfo {pages} {105} (\bibinfo {year} {2015})}\BibitemShut
  {NoStop}%
\bibitem [{\citenamefont {Carrillo}\ \emph {et~al.}(2018)\citenamefont
  {Carrillo}, \citenamefont {Minchev}, \citenamefont {Kordopatis},
  \citenamefont {Steinmetz}, \citenamefont {Binney}, \citenamefont {Anders},
  \citenamefont {Bienaym\'{e}}, \citenamefont {Bland-Hawthorn}, \citenamefont
  {Famaey}, \citenamefont {Freeman} \emph {et~al.}}]{Carrillo2018}%
  \BibitemOpen
  \bibfield  {author} {\bibinfo {author} {\bibfnamefont {I.}~\bibnamefont
  {Carrillo}}, \bibinfo {author} {\bibfnamefont {I.}~\bibnamefont {Minchev}},
  \bibinfo {author} {\bibfnamefont {G.}~\bibnamefont {Kordopatis}}, \bibinfo
  {author} {\bibfnamefont {M.}~\bibnamefont {Steinmetz}}, \bibinfo {author}
  {\bibfnamefont {J.}~\bibnamefont {Binney}}, \bibinfo {author} {\bibfnamefont
  {F.}~\bibnamefont {Anders}}, \bibinfo {author} {\bibfnamefont
  {O.}~\bibnamefont {Bienaym\'{e}}}, \bibinfo {author} {\bibfnamefont
  {J.}~\bibnamefont {Bland-Hawthorn}}, \bibinfo {author} {\bibfnamefont
  {B.}~\bibnamefont {Famaey}}, \bibinfo {author} {\bibfnamefont {K.~C.}\
  \bibnamefont {Freeman}},  \emph {et~al.},\ }\href {\doibase gc8mwq}
  {\bibfield  {journal} {\bibinfo  {journal} {Mon. Not. R. Astron. Soc.}\
  }\textbf {\bibinfo {volume} {475}},\ \bibinfo {pages} {2679} (\bibinfo {year}
  {2018})}\BibitemShut {NoStop}%
\bibitem [{\citenamefont {Sch\"{o}nrich}\ and\ \citenamefont
  {Dehnen}(2018)}]{Schoenrich2018}%
  \BibitemOpen
  \bibfield  {author} {\bibinfo {author} {\bibfnamefont {R.}~\bibnamefont
  {Sch\"{o}nrich}}\ and\ \bibinfo {author} {\bibfnamefont {W.}~\bibnamefont
  {Dehnen}},\ }\href {\doibase 10.1093/mnras/sty1256} {\bibfield  {journal}
  {\bibinfo  {journal} {Mon. Not. R. Astron. Soc.}\ }\textbf {\bibinfo {volume}
  {478}},\ \bibinfo {pages} {3809} (\bibinfo {year} {2018})}\BibitemShut
  {NoStop}%
\bibitem [{\citenamefont {Antoja}\ \emph {et~al.}(2018)\citenamefont {Antoja},
  \citenamefont {Helmi}, \citenamefont {Romero-G\'{o}mez}, \citenamefont
  {Katz}, \citenamefont {Babusiaux}, \citenamefont {Drimmel}, \citenamefont
  {Evans}, \citenamefont {Figueras}, \citenamefont {Poggio}, \citenamefont
  {Reyl\'{e}} \emph {et~al.}}]{Antoja2018}%
  \BibitemOpen
  \bibfield  {author} {\bibinfo {author} {\bibfnamefont {T.}~\bibnamefont
  {Antoja}}, \bibinfo {author} {\bibfnamefont {A.}~\bibnamefont {Helmi}},
  \bibinfo {author} {\bibfnamefont {M.}~\bibnamefont {Romero-G\'{o}mez}},
  \bibinfo {author} {\bibfnamefont {D.}~\bibnamefont {Katz}}, \bibinfo {author}
  {\bibfnamefont {C.}~\bibnamefont {Babusiaux}}, \bibinfo {author}
  {\bibfnamefont {R.}~\bibnamefont {Drimmel}}, \bibinfo {author} {\bibfnamefont
  {D.~W.}\ \bibnamefont {Evans}}, \bibinfo {author} {\bibfnamefont
  {F.}~\bibnamefont {Figueras}}, \bibinfo {author} {\bibfnamefont
  {E.}~\bibnamefont {Poggio}}, \bibinfo {author} {\bibfnamefont
  {C.}~\bibnamefont {Reyl\'{e}}},  \emph {et~al.},\ }\href {\doibase gd7gh3}
  {\bibfield  {journal} {\bibinfo  {journal} {Nature}\ }\textbf {\bibinfo
  {volume} {561}},\ \bibinfo {pages} {360} (\bibinfo {year}
  {2018})}\BibitemShut {NoStop}%
\bibitem [{\citenamefont {Laporte}\ \emph {et~al.}(2019)\citenamefont
  {Laporte}, \citenamefont {Minchev}, \citenamefont {Johnston},\ and\
  \citenamefont {G\'{o}mez}}]{Laporte2019}%
  \BibitemOpen
  \bibfield  {author} {\bibinfo {author} {\bibfnamefont {C.~F.~P.}\
  \bibnamefont {Laporte}}, \bibinfo {author} {\bibfnamefont {I.}~\bibnamefont
  {Minchev}}, \bibinfo {author} {\bibfnamefont {K.~V.}\ \bibnamefont
  {Johnston}}, \ and\ \bibinfo {author} {\bibfnamefont {F.~A.}\ \bibnamefont
  {G\'{o}mez}},\ }\href {\doibase 10.1093/mnras/stz583} {\bibfield  {journal}
  {\bibinfo  {journal} {Mon. Not. R. Astron. Soc.}\ }\textbf {\bibinfo {volume}
  {485}},\ \bibinfo {pages} {3134} (\bibinfo {year} {2019})}\BibitemShut
  {NoStop}%
\bibitem [{\citenamefont {{Bennett}}\ and\ \citenamefont
  {{Bovy}}(2019)}]{Bennett2019}%
  \BibitemOpen
  \bibfield  {author} {\bibinfo {author} {\bibfnamefont {M.}~\bibnamefont
  {{Bennett}}}\ and\ \bibinfo {author} {\bibfnamefont {J.}~\bibnamefont
  {{Bovy}}},\ }\href {\doibase 10.1093/mnras/sty2813} {\bibfield  {journal}
  {\bibinfo  {journal} {Mon. Not. R. Astron. Soc.}\ }\textbf {\bibinfo {volume}
  {482}},\ \bibinfo {pages} {1417} (\bibinfo {year} {2019})}\BibitemShut
  {NoStop}%
\bibitem [{\citenamefont {Necib}\ \emph
  {et~al.}(2019{\natexlab{a}})\citenamefont {Necib}, \citenamefont {Lisanti},\
  and\ \citenamefont {Belokurov}}]{Necib2019}%
  \BibitemOpen
  \bibfield  {author} {\bibinfo {author} {\bibfnamefont {L.}~\bibnamefont
  {Necib}}, \bibinfo {author} {\bibfnamefont {M.}~\bibnamefont {Lisanti}}, \
  and\ \bibinfo {author} {\bibfnamefont {V.}~\bibnamefont {Belokurov}},\ }\href
  {\doibase 10.3847/1538-4357/ab095b} {\bibfield  {journal} {\bibinfo
  {journal} {Astrophys. J.}\ }\textbf {\bibinfo {volume} {874}},\ \bibinfo
  {pages} {3} (\bibinfo {year} {2019}{\natexlab{a}})}\BibitemShut {NoStop}%
\bibitem [{\citenamefont {Necib}\ \emph
  {et~al.}(2019{\natexlab{b}})\citenamefont {Necib}, \citenamefont {Ostdiek},
  \citenamefont {Lisanti}, \citenamefont {Cohen}, \citenamefont {Freytsis},\
  and\ \citenamefont {Garrison-Kimmel}}]{Necib2019a}%
  \BibitemOpen
  \bibfield  {author} {\bibinfo {author} {\bibfnamefont {L.}~\bibnamefont
  {Necib}}, \bibinfo {author} {\bibfnamefont {B.}~\bibnamefont {Ostdiek}},
  \bibinfo {author} {\bibfnamefont {M.}~\bibnamefont {Lisanti}}, \bibinfo
  {author} {\bibfnamefont {T.}~\bibnamefont {Cohen}}, \bibinfo {author}
  {\bibfnamefont {M.}~\bibnamefont {Freytsis}}, \ and\ \bibinfo {author}
  {\bibfnamefont {S.}~\bibnamefont {Garrison-Kimmel}},\ }\href@noop {} {}
  (\bibinfo {year} {2019}{\natexlab{b}}),\ \Eprint
  {http://arxiv.org/abs/1907.07681} {arXiv:1907.07681 [astro-ph.GA]}
  \BibitemShut {NoStop}%
\bibitem [{\citenamefont {Hartnett}\ and\ \citenamefont
  {Luiten}(2011)}]{Hartnett2011}%
  \BibitemOpen
  \bibfield  {author} {\bibinfo {author} {\bibfnamefont {J.~G.}\ \bibnamefont
  {Hartnett}}\ and\ \bibinfo {author} {\bibfnamefont {A.~N.}\ \bibnamefont
  {Luiten}},\ }\href {\doibase 10.1103/RevModPhys.83.1} {\bibfield  {journal}
  {\bibinfo  {journal} {Rev. Mod. Phys.}\ }\textbf {\bibinfo {volume} {83}},\
  \bibinfo {pages} {1} (\bibinfo {year} {2011})}\BibitemShut {NoStop}%
\bibitem [{\citenamefont {Manchester}(2017)}]{Manchester2017}%
  \BibitemOpen
  \bibfield  {author} {\bibinfo {author} {\bibfnamefont {R.~N.}\ \bibnamefont
  {Manchester}},\ }\href {\doibase 10.1007/s12036-017-9469-2} {\bibfield
  {journal} {\bibinfo  {journal} {J. Astrophys. Astron.}\ }\textbf {\bibinfo
  {volume} {38}},\ \bibinfo {pages} {42} (\bibinfo {year} {2017})}\BibitemShut
  {NoStop}%
\bibitem [{\citenamefont {Weisberg}\ \emph {et~al.}(2010)\citenamefont
  {Weisberg}, \citenamefont {Nice},\ and\ \citenamefont
  {Taylor}}]{Weisberg2010}%
  \BibitemOpen
  \bibfield  {author} {\bibinfo {author} {\bibfnamefont {J.~M.}\ \bibnamefont
  {Weisberg}}, \bibinfo {author} {\bibfnamefont {D.~J.}\ \bibnamefont {Nice}},
  \ and\ \bibinfo {author} {\bibfnamefont {J.~H.}\ \bibnamefont {Taylor}},\
  }\href {\doibase 10.1088/0004-637X/722/2/1030} {\bibfield  {journal}
  {\bibinfo  {journal} {Astrophys. J.}\ }\textbf {\bibinfo {volume} {722}},\
  \bibinfo {pages} {1030} (\bibinfo {year} {2010})}\BibitemShut {NoStop}%
\bibitem [{\citenamefont {Weisberg}\ and\ \citenamefont
  {Huang}(2016)}]{Weisberg2016}%
  \BibitemOpen
  \bibfield  {author} {\bibinfo {author} {\bibfnamefont {J.~M.}\ \bibnamefont
  {Weisberg}}\ and\ \bibinfo {author} {\bibfnamefont {Y.}~\bibnamefont
  {Huang}},\ }\href {\doibase 10.3847/0004-637X/829/1/55} {\bibfield  {journal}
  {\bibinfo  {journal} {Astrophys. J.}\ }\textbf {\bibinfo {volume} {829}},\
  \bibinfo {pages} {55} (\bibinfo {year} {2016})}\BibitemShut {NoStop}%
\bibitem [{\citenamefont {Ransom}\ \emph {et~al.}(2014)\citenamefont {Ransom},
  \citenamefont {Stairs}, \citenamefont {Archibald}, \citenamefont {Hessels},
  \citenamefont {Kaplan}, \citenamefont {van Kerkwijk}, \citenamefont {Boyles},
  \citenamefont {Deller}, \citenamefont {Chatterjee}, \citenamefont
  {Schechtman-Rook} \emph {et~al.}}]{Ransom2014}%
  \BibitemOpen
  \bibfield  {author} {\bibinfo {author} {\bibfnamefont {S.~M.}\ \bibnamefont
  {Ransom}}, \bibinfo {author} {\bibfnamefont {I.~H.}\ \bibnamefont {Stairs}},
  \bibinfo {author} {\bibfnamefont {A.~M.}\ \bibnamefont {Archibald}}, \bibinfo
  {author} {\bibfnamefont {J.~W.~T.}\ \bibnamefont {Hessels}}, \bibinfo
  {author} {\bibfnamefont {D.~L.}\ \bibnamefont {Kaplan}}, \bibinfo {author}
  {\bibfnamefont {M.~H.}\ \bibnamefont {van Kerkwijk}}, \bibinfo {author}
  {\bibfnamefont {J.}~\bibnamefont {Boyles}}, \bibinfo {author} {\bibfnamefont
  {A.~T.}\ \bibnamefont {Deller}}, \bibinfo {author} {\bibfnamefont
  {S.}~\bibnamefont {Chatterjee}}, \bibinfo {author} {\bibfnamefont
  {A.}~\bibnamefont {Schechtman-Rook}},  \emph {et~al.},\ }\href {\doibase
  f5nxqj} {\bibfield  {journal} {\bibinfo  {journal} {Nature}\ }\textbf
  {\bibinfo {volume} {505}},\ \bibinfo {pages} {520} (\bibinfo {year}
  {2014})}\BibitemShut {NoStop}%
\bibitem [{\citenamefont {Archibald}\ \emph {et~al.}(2018)\citenamefont
  {Archibald}, \citenamefont {Gusinskaia}, \citenamefont {Hessels},
  \citenamefont {Deller}, \citenamefont {Kaplan}, \citenamefont {Lorimer},
  \citenamefont {Lynch}, \citenamefont {Ransom},\ and\ \citenamefont
  {Stairs}}]{Archibald2018}%
  \BibitemOpen
  \bibfield  {author} {\bibinfo {author} {\bibfnamefont {A.~M.}\ \bibnamefont
  {Archibald}}, \bibinfo {author} {\bibfnamefont {N.~V.}\ \bibnamefont
  {Gusinskaia}}, \bibinfo {author} {\bibfnamefont {J.~W.~T.}\ \bibnamefont
  {Hessels}}, \bibinfo {author} {\bibfnamefont {A.~T.}\ \bibnamefont {Deller}},
  \bibinfo {author} {\bibfnamefont {D.~L.}\ \bibnamefont {Kaplan}}, \bibinfo
  {author} {\bibfnamefont {D.~R.}\ \bibnamefont {Lorimer}}, \bibinfo {author}
  {\bibfnamefont {R.~S.}\ \bibnamefont {Lynch}}, \bibinfo {author}
  {\bibfnamefont {S.~M.}\ \bibnamefont {Ransom}}, \ and\ \bibinfo {author}
  {\bibfnamefont {I.~H.}\ \bibnamefont {Stairs}},\ }\href {\doibase crqp}
  {\bibfield  {journal} {\bibinfo  {journal} {Nature}\ }\textbf {\bibinfo
  {volume} {559}},\ \bibinfo {pages} {73} (\bibinfo {year} {2018})}\BibitemShut
  {NoStop}%
\bibitem [{\citenamefont {Voisin}\ \emph {et~al.}(2020)\citenamefont {Voisin},
  \citenamefont {Cognard}, \citenamefont {Freire}, \citenamefont {Wex},
  \citenamefont {Guillemot}, \citenamefont {Desvignes}, \citenamefont
  {Kramer},\ and\ \citenamefont {Theureau}}]{Voisin2020}%
  \BibitemOpen
  \bibfield  {author} {\bibinfo {author} {\bibfnamefont {G.}~\bibnamefont
  {Voisin}}, \bibinfo {author} {\bibfnamefont {I.}~\bibnamefont {Cognard}},
  \bibinfo {author} {\bibfnamefont {P.~C.~C.}\ \bibnamefont {Freire}}, \bibinfo
  {author} {\bibfnamefont {N.}~\bibnamefont {Wex}}, \bibinfo {author}
  {\bibfnamefont {L.}~\bibnamefont {Guillemot}}, \bibinfo {author}
  {\bibfnamefont {G.}~\bibnamefont {Desvignes}}, \bibinfo {author}
  {\bibfnamefont {M.}~\bibnamefont {Kramer}}, \ and\ \bibinfo {author}
  {\bibfnamefont {G.}~\bibnamefont {Theureau}},\ }\href {\doibase dzcf}
  {\bibfield  {journal} {\bibinfo  {journal} {Astron. Astrophys.}\ }\textbf
  {\bibinfo {volume} {638}},\ \bibinfo {pages} {A24} (\bibinfo {year}
  {2020})}\BibitemShut {NoStop}%
\bibitem [{\citenamefont {Lommen}(2015)}]{Lommen2015}%
  \BibitemOpen
  \bibfield  {author} {\bibinfo {author} {\bibfnamefont {A.~N.}\ \bibnamefont
  {Lommen}},\ }\href {\doibase 10.1088/0034-4885/78/12/124901} {\bibfield
  {journal} {\bibinfo  {journal} {Rep. Prog. Phys.}\ }\textbf {\bibinfo
  {volume} {78}},\ \bibinfo {pages} {124901} (\bibinfo {year}
  {2015})}\BibitemShut {NoStop}%
\bibitem [{\citenamefont {Shannon}\ \emph {et~al.}(2013)\citenamefont
  {Shannon}, \citenamefont {Ravi}, \citenamefont {Coles}, \citenamefont
  {Hobbs}, \citenamefont {Keith}, \citenamefont {Manchester}, \citenamefont
  {Wyithe}, \citenamefont {Bailes}, \citenamefont {Bhat}, \citenamefont
  {Burke-Spolaor} \emph {et~al.}}]{Shannon2013}%
  \BibitemOpen
  \bibfield  {author} {\bibinfo {author} {\bibfnamefont {R.~M.}\ \bibnamefont
  {Shannon}}, \bibinfo {author} {\bibfnamefont {V.}~\bibnamefont {Ravi}},
  \bibinfo {author} {\bibfnamefont {W.~A.}\ \bibnamefont {Coles}}, \bibinfo
  {author} {\bibfnamefont {G.}~\bibnamefont {Hobbs}}, \bibinfo {author}
  {\bibfnamefont {M.~J.}\ \bibnamefont {Keith}}, \bibinfo {author}
  {\bibfnamefont {R.~N.}\ \bibnamefont {Manchester}}, \bibinfo {author}
  {\bibfnamefont {J.~S.~B.}\ \bibnamefont {Wyithe}}, \bibinfo {author}
  {\bibfnamefont {M.}~\bibnamefont {Bailes}}, \bibinfo {author} {\bibfnamefont
  {N.~D.~R.}\ \bibnamefont {Bhat}}, \bibinfo {author} {\bibfnamefont
  {S.}~\bibnamefont {Burke-Spolaor}},  \emph {et~al.},\ }\href
  {http://adsabs.harvard.edu/abs/2013Sci...342..334S} {\bibfield  {journal}
  {\bibinfo  {journal} {Science}\ }\textbf {\bibinfo {volume} {342}},\ \bibinfo
  {pages} {334} (\bibinfo {year} {2013})}\BibitemShut {NoStop}%
\bibitem [{\citenamefont {Shannon}\ \emph {et~al.}(2015)\citenamefont
  {Shannon}, \citenamefont {Ravi}, \citenamefont {Lentati}, \citenamefont
  {Lasky}, \citenamefont {Hobbs}, \citenamefont {Kerr}, \citenamefont
  {Manchester}, \citenamefont {Coles}, \citenamefont {Levin}, \citenamefont
  {Bailes} \emph {et~al.}}]{Shannon2015}%
  \BibitemOpen
  \bibfield  {author} {\bibinfo {author} {\bibfnamefont {R.~M.}\ \bibnamefont
  {Shannon}}, \bibinfo {author} {\bibfnamefont {V.}~\bibnamefont {Ravi}},
  \bibinfo {author} {\bibfnamefont {L.~T.}\ \bibnamefont {Lentati}}, \bibinfo
  {author} {\bibfnamefont {P.~D.}\ \bibnamefont {Lasky}}, \bibinfo {author}
  {\bibfnamefont {G.}~\bibnamefont {Hobbs}}, \bibinfo {author} {\bibfnamefont
  {M.}~\bibnamefont {Kerr}}, \bibinfo {author} {\bibfnamefont {R.~N.}\
  \bibnamefont {Manchester}}, \bibinfo {author} {\bibfnamefont {W.~A.}\
  \bibnamefont {Coles}}, \bibinfo {author} {\bibfnamefont {Y.}~\bibnamefont
  {Levin}}, \bibinfo {author} {\bibfnamefont {M.}~\bibnamefont {Bailes}},
  \emph {et~al.},\ }\href {\doibase 10.1126/science.aab1910} {\bibfield
  {journal} {\bibinfo  {journal} {Science}\ }\textbf {\bibinfo {volume}
  {349}},\ \bibinfo {pages} {1522} (\bibinfo {year} {2015})}\BibitemShut
  {NoStop}%
\bibitem [{\citenamefont {Demorest}\ \emph {et~al.}(2013)\citenamefont
  {Demorest}, \citenamefont {Ferdman}, \citenamefont {Gonzalez}, \citenamefont
  {Nice}, \citenamefont {Ransom}, \citenamefont {Stairs}, \citenamefont
  {Arzoumanian}, \citenamefont {Brazier}, \citenamefont {Burke-Spolaor},
  \citenamefont {Chamberlin} \emph {et~al.}}]{Demorest2013}%
  \BibitemOpen
  \bibfield  {author} {\bibinfo {author} {\bibfnamefont {P.~B.}\ \bibnamefont
  {Demorest}}, \bibinfo {author} {\bibfnamefont {R.~D.}\ \bibnamefont
  {Ferdman}}, \bibinfo {author} {\bibfnamefont {M.~E.}\ \bibnamefont
  {Gonzalez}}, \bibinfo {author} {\bibfnamefont {D.}~\bibnamefont {Nice}},
  \bibinfo {author} {\bibfnamefont {S.}~\bibnamefont {Ransom}}, \bibinfo
  {author} {\bibfnamefont {I.~H.}\ \bibnamefont {Stairs}}, \bibinfo {author}
  {\bibfnamefont {Z.}~\bibnamefont {Arzoumanian}}, \bibinfo {author}
  {\bibfnamefont {A.}~\bibnamefont {Brazier}}, \bibinfo {author} {\bibfnamefont
  {S.}~\bibnamefont {Burke-Spolaor}}, \bibinfo {author} {\bibfnamefont {S.~J.}\
  \bibnamefont {Chamberlin}},  \emph {et~al.},\ }\href
  {https://iopscience.iop.org/article/10.1088/0004-637X/762/2/94} {\bibfield
  {journal} {\bibinfo  {journal} {Astrophys. J.}\ }\textbf {\bibinfo {volume}
  {762}},\ \bibinfo {pages} {94} (\bibinfo {year} {2013})}\BibitemShut
  {NoStop}%
\bibitem [{\citenamefont {Lentati}\ \emph {et~al.}(2015)\citenamefont
  {Lentati}, \citenamefont {Taylor}, \citenamefont {Mingarelli}, \citenamefont
  {Sesana}, \citenamefont {Sanidas}, \citenamefont {Vecchio}, \citenamefont
  {Caballero}, \citenamefont {Lee}, \citenamefont {van Haasteren},
  \citenamefont {Babak} \emph {et~al.}}]{Lentati2015}%
  \BibitemOpen
  \bibfield  {author} {\bibinfo {author} {\bibfnamefont {L.}~\bibnamefont
  {Lentati}}, \bibinfo {author} {\bibfnamefont {S.~R.}\ \bibnamefont {Taylor}},
  \bibinfo {author} {\bibfnamefont {C.~M.~F.}\ \bibnamefont {Mingarelli}},
  \bibinfo {author} {\bibfnamefont {A.}~\bibnamefont {Sesana}}, \bibinfo
  {author} {\bibfnamefont {S.~A.}\ \bibnamefont {Sanidas}}, \bibinfo {author}
  {\bibfnamefont {A.}~\bibnamefont {Vecchio}}, \bibinfo {author} {\bibfnamefont
  {R.~N.}\ \bibnamefont {Caballero}}, \bibinfo {author} {\bibfnamefont {K.~J.}\
  \bibnamefont {Lee}}, \bibinfo {author} {\bibfnamefont {R.}~\bibnamefont {van
  Haasteren}}, \bibinfo {author} {\bibfnamefont {S.}~\bibnamefont {Babak}},
  \emph {et~al.},\ }\href {\doibase 10.1093/mnras/stv1538} {\bibfield
  {journal} {\bibinfo  {journal} {Mon. Not. R. Astron. Soc.}\ }\textbf
  {\bibinfo {volume} {453}},\ \bibinfo {pages} {2576} (\bibinfo {year}
  {2015})}\BibitemShut {NoStop}%
\bibitem [{\citenamefont {Babak}\ \emph {et~al.}(2016)\citenamefont {Babak},
  \citenamefont {Petiteau}, \citenamefont {Sesana}, \citenamefont {Brem},
  \citenamefont {Rosado}, \citenamefont {Taylor}, \citenamefont {Lassus},
  \citenamefont {Hessels}, \citenamefont {Bassa}, \citenamefont {Burgay} \emph
  {et~al.}}]{Babak2016}%
  \BibitemOpen
  \bibfield  {author} {\bibinfo {author} {\bibfnamefont {S.}~\bibnamefont
  {Babak}}, \bibinfo {author} {\bibfnamefont {A.}~\bibnamefont {Petiteau}},
  \bibinfo {author} {\bibfnamefont {A.}~\bibnamefont {Sesana}}, \bibinfo
  {author} {\bibfnamefont {P.}~\bibnamefont {Brem}}, \bibinfo {author}
  {\bibfnamefont {P.~A.}\ \bibnamefont {Rosado}}, \bibinfo {author}
  {\bibfnamefont {S.~R.}\ \bibnamefont {Taylor}}, \bibinfo {author}
  {\bibfnamefont {A.}~\bibnamefont {Lassus}}, \bibinfo {author} {\bibfnamefont
  {J.~W.~T.}\ \bibnamefont {Hessels}}, \bibinfo {author} {\bibfnamefont
  {C.~G.}\ \bibnamefont {Bassa}}, \bibinfo {author} {\bibfnamefont
  {M.}~\bibnamefont {Burgay}},  \emph {et~al.},\ }\href {\doibase f7636b}
  {\bibfield  {journal} {\bibinfo  {journal} {Mon. Not. R. Astron. Soc.}\
  }\textbf {\bibinfo {volume} {455}},\ \bibinfo {pages} {1665} (\bibinfo {year}
  {2016})}\BibitemShut {NoStop}%
\bibitem [{\citenamefont {Arzoumanian}\ \emph {et~al.}(2016)\citenamefont
  {Arzoumanian}, \citenamefont {Brazier}, \citenamefont {Burke-Spolaor},
  \citenamefont {Chamberlin}, \citenamefont {Chatterjee}, \citenamefont
  {Christy}, \citenamefont {Cordes}, \citenamefont {Cornish}, \citenamefont
  {Crowter}, \citenamefont {Demorest} \emph {et~al.}}]{Arzoumanian2016}%
  \BibitemOpen
  \bibfield  {author} {\bibinfo {author} {\bibfnamefont {Z.}~\bibnamefont
  {Arzoumanian}}, \bibinfo {author} {\bibfnamefont {A.}~\bibnamefont
  {Brazier}}, \bibinfo {author} {\bibfnamefont {S.}~\bibnamefont
  {Burke-Spolaor}}, \bibinfo {author} {\bibfnamefont {S.~J.}\ \bibnamefont
  {Chamberlin}}, \bibinfo {author} {\bibfnamefont {S.}~\bibnamefont
  {Chatterjee}}, \bibinfo {author} {\bibfnamefont {B.}~\bibnamefont {Christy}},
  \bibinfo {author} {\bibfnamefont {J.~M.}\ \bibnamefont {Cordes}}, \bibinfo
  {author} {\bibfnamefont {N.~J.}\ \bibnamefont {Cornish}}, \bibinfo {author}
  {\bibfnamefont {K.}~\bibnamefont {Crowter}}, \bibinfo {author} {\bibfnamefont
  {P.~B.}\ \bibnamefont {Demorest}},  \emph {et~al.},\ }\href {\doibase ggtpwd}
  {\bibfield  {journal} {\bibinfo  {journal} {Astrophys. J.}\ }\textbf
  {\bibinfo {volume} {821}},\ \bibinfo {pages} {13} (\bibinfo {year}
  {2016})}\BibitemShut {NoStop}%
\bibitem [{\citenamefont {Aggarwal}\ \emph {et~al.}(2019)\citenamefont
  {Aggarwal}, \citenamefont {Arzoumanian}, \citenamefont {Baker}, \citenamefont
  {Brazier}, \citenamefont {Brinson}, \citenamefont {Brook}, \citenamefont
  {Burke-Spolaor}, \citenamefont {Chatterjee}, \citenamefont {Cordes},
  \citenamefont {Cornish} \emph {et~al.}}]{Aggarwal2019}%
  \BibitemOpen
  \bibfield  {author} {\bibinfo {author} {\bibfnamefont {K.}~\bibnamefont
  {Aggarwal}}, \bibinfo {author} {\bibfnamefont {Z.}~\bibnamefont
  {Arzoumanian}}, \bibinfo {author} {\bibfnamefont {P.~T.}\ \bibnamefont
  {Baker}}, \bibinfo {author} {\bibfnamefont {A.}~\bibnamefont {Brazier}},
  \bibinfo {author} {\bibfnamefont {M.~R.}\ \bibnamefont {Brinson}}, \bibinfo
  {author} {\bibfnamefont {P.~R.}\ \bibnamefont {Brook}}, \bibinfo {author}
  {\bibfnamefont {S.}~\bibnamefont {Burke-Spolaor}}, \bibinfo {author}
  {\bibfnamefont {S.}~\bibnamefont {Chatterjee}}, \bibinfo {author}
  {\bibfnamefont {J.~M.}\ \bibnamefont {Cordes}}, \bibinfo {author}
  {\bibfnamefont {N.~J.}\ \bibnamefont {Cornish}},  \emph {et~al.},\ }\href
  {\doibase 10.3847/1538-4357/ab2236} {\bibfield  {journal} {\bibinfo
  {journal} {Astrophys. J.}\ }\textbf {\bibinfo {volume} {880}},\ \bibinfo
  {pages} {116} (\bibinfo {year} {2019})}\BibitemShut {NoStop}%
\bibitem [{\citenamefont {Verbiest}\ \emph {et~al.}(2016)\citenamefont
  {Verbiest}, \citenamefont {Lentati}, \citenamefont {Hobbs}, \citenamefont
  {van Haasteren}, \citenamefont {Demorest}, \citenamefont {Janssen},
  \citenamefont {Wang}, \citenamefont {Desvignes}, \citenamefont {Caballero},
  \citenamefont {Keith} \emph {et~al.}}]{Verbiest2016}%
  \BibitemOpen
  \bibfield  {author} {\bibinfo {author} {\bibfnamefont {J.~P.~W.}\
  \bibnamefont {Verbiest}}, \bibinfo {author} {\bibfnamefont {L.}~\bibnamefont
  {Lentati}}, \bibinfo {author} {\bibfnamefont {G.}~\bibnamefont {Hobbs}},
  \bibinfo {author} {\bibfnamefont {R.}~\bibnamefont {van Haasteren}}, \bibinfo
  {author} {\bibfnamefont {P.~B.}\ \bibnamefont {Demorest}}, \bibinfo {author}
  {\bibfnamefont {G.~H.}\ \bibnamefont {Janssen}}, \bibinfo {author}
  {\bibfnamefont {J.-B.}\ \bibnamefont {Wang}}, \bibinfo {author}
  {\bibfnamefont {G.}~\bibnamefont {Desvignes}}, \bibinfo {author}
  {\bibfnamefont {R.~N.}\ \bibnamefont {Caballero}}, \bibinfo {author}
  {\bibfnamefont {M.~J.}\ \bibnamefont {Keith}},  \emph {et~al.},\ }\href
  {\doibase 10.1093/mnras/stw347} {\bibfield  {journal} {\bibinfo  {journal}
  {Mon. Not. R. Astron. Soc.}\ }\textbf {\bibinfo {volume} {458}},\ \bibinfo
  {pages} {1267} (\bibinfo {year} {2016})}\BibitemShut {NoStop}%
\bibitem [{\citenamefont {Dubovsky}\ \emph {et~al.}(2005)\citenamefont
  {Dubovsky}, \citenamefont {Tinyakov},\ and\ \citenamefont
  {Tkachev}}]{Dubovsky2005}%
  \BibitemOpen
  \bibfield  {author} {\bibinfo {author} {\bibfnamefont {S.}~\bibnamefont
  {Dubovsky}}, \bibinfo {author} {\bibfnamefont {P.}~\bibnamefont {Tinyakov}},
  \ and\ \bibinfo {author} {\bibfnamefont {I.}~\bibnamefont {Tkachev}},\ }\href
  {\doibase bn4jkz} {\bibfield  {journal} {\bibinfo  {journal} {Phys. Rev.
  Lett.}\ }\textbf {\bibinfo {volume} {{94}}},\ \bibinfo {pages} {181102}
  (\bibinfo {year} {{2005}})}\BibitemShut {NoStop}%
\bibitem [{\citenamefont {Baskaran}\ \emph {et~al.}(2008)\citenamefont
  {Baskaran}, \citenamefont {Polnarev}, \citenamefont {Pshirkov},\ and\
  \citenamefont {Postnov}}]{Baskaran2008}%
  \BibitemOpen
  \bibfield  {author} {\bibinfo {author} {\bibfnamefont {D.}~\bibnamefont
  {Baskaran}}, \bibinfo {author} {\bibfnamefont {A.~G.}\ \bibnamefont
  {Polnarev}}, \bibinfo {author} {\bibfnamefont {M.~S.}\ \bibnamefont
  {Pshirkov}}, \ and\ \bibinfo {author} {\bibfnamefont {K.~A.}\ \bibnamefont
  {Postnov}},\ }\href {\doibase ddsdx3} {\bibfield  {journal} {\bibinfo
  {journal} {Phys. Rev. D}\ }\textbf {\bibinfo {volume} {{78}}},\ \bibinfo
  {pages} {044018} (\bibinfo {year} {{2008}})}\BibitemShut {NoStop}%
\bibitem [{\citenamefont {Pshirkov}\ \emph {et~al.}(2008)\citenamefont
  {Pshirkov}, \citenamefont {Tuntsov},\ and\ \citenamefont
  {Postnov}}]{Pshirkov2008}%
  \BibitemOpen
  \bibfield  {author} {\bibinfo {author} {\bibfnamefont {M.}~\bibnamefont
  {Pshirkov}}, \bibinfo {author} {\bibfnamefont {A.}~\bibnamefont {Tuntsov}}, \
  and\ \bibinfo {author} {\bibfnamefont {K.~A.}\ \bibnamefont {Postnov}},\
  }\href {\doibase bfxqfs} {\bibfield  {journal} {\bibinfo  {journal} {Phys.
  Rev. Lett.}\ }\textbf {\bibinfo {volume} {{101}}},\ \bibinfo {pages} {261101}
  (\bibinfo {year} {{2008}})}\BibitemShut {NoStop}%
\bibitem [{\citenamefont {Seto}\ and\ \citenamefont {Cooray}(2007)}]{Seto2007}%
  \BibitemOpen
  \bibfield  {author} {\bibinfo {author} {\bibfnamefont {N.}~\bibnamefont
  {Seto}}\ and\ \bibinfo {author} {\bibfnamefont {A.}~\bibnamefont {Cooray}},\
  }\href {\doibase 10.1086/516570} {\bibfield  {journal} {\bibinfo  {journal}
  {Astrophys. J.}\ }\textbf {\bibinfo {volume} {659}},\ \bibinfo {pages} {L33}
  (\bibinfo {year} {2007})}\BibitemShut {NoStop}%
\bibitem [{\citenamefont {Kashiyama}\ and\ \citenamefont
  {Seto}(2012)}]{Kashiyama2012}%
  \BibitemOpen
  \bibfield  {author} {\bibinfo {author} {\bibfnamefont {K.}~\bibnamefont
  {Kashiyama}}\ and\ \bibinfo {author} {\bibfnamefont {N.}~\bibnamefont
  {Seto}},\ }\href {\doibase 10.1111/j.1365-2966.2012.21935.x} {\bibfield
  {journal} {\bibinfo  {journal} {Mon. Not. R. Astron. Soc.}\ }\textbf
  {\bibinfo {volume} {426}},\ \bibinfo {pages} {1369} (\bibinfo {year}
  {2012})}\BibitemShut {NoStop}%
\bibitem [{\citenamefont {Schutz}\ and\ \citenamefont
  {Liu}(2017)}]{Schutz2017}%
  \BibitemOpen
  \bibfield  {author} {\bibinfo {author} {\bibfnamefont {K.}~\bibnamefont
  {Schutz}}\ and\ \bibinfo {author} {\bibfnamefont {A.}~\bibnamefont {Liu}},\
  }\href {\doibase 10.1103/PhysRevD.95.023002} {\bibfield  {journal} {\bibinfo
  {journal} {Phys. Rev. D}\ }\textbf {\bibinfo {volume} {95}},\ \bibinfo
  {pages} {023002} (\bibinfo {year} {2017})}\BibitemShut {NoStop}%
\bibitem [{\citenamefont {Dror}\ \emph {et~al.}(2019)\citenamefont {Dror},
  \citenamefont {Ramani}, \citenamefont {Trickle},\ and\ \citenamefont
  {Zurek}}]{Dror2019}%
  \BibitemOpen
  \bibfield  {author} {\bibinfo {author} {\bibfnamefont {J.~A.}\ \bibnamefont
  {Dror}}, \bibinfo {author} {\bibfnamefont {H.}~\bibnamefont {Ramani}},
  \bibinfo {author} {\bibfnamefont {T.}~\bibnamefont {Trickle}}, \ and\
  \bibinfo {author} {\bibfnamefont {K.~M.}\ \bibnamefont {Zurek}},\ }\href
  {\doibase 10.1103/PhysRevD.100.023003} {\bibfield  {journal} {\bibinfo
  {journal} {Phys. Rev. D}\ }\textbf {\bibinfo {volume} {100}},\ \bibinfo
  {pages} {023003} (\bibinfo {year} {2019})}\BibitemShut {NoStop}%
\bibitem [{\citenamefont {Khmelnitsky}\ and\ \citenamefont
  {Rubakov}(2014)}]{Khmelnitsky2014}%
  \BibitemOpen
  \bibfield  {author} {\bibinfo {author} {\bibfnamefont {A.}~\bibnamefont
  {Khmelnitsky}}\ and\ \bibinfo {author} {\bibfnamefont {V.}~\bibnamefont
  {Rubakov}},\ }\href {\doibase 10.1088/1475-7516/2014/02/019} {\bibfield
  {journal} {\bibinfo  {journal} {J. Cosmol. Astropart. Phys.}\ }\textbf
  {\bibinfo {volume} {2014}},\ \bibinfo {pages} {019} (\bibinfo {year}
  {{2014}})}\BibitemShut {NoStop}%
\bibitem [{\citenamefont {Porayko}\ and\ \citenamefont
  {Postnov}(2014)}]{Porayko2014}%
  \BibitemOpen
  \bibfield  {author} {\bibinfo {author} {\bibfnamefont {N.~K.}\ \bibnamefont
  {Porayko}}\ and\ \bibinfo {author} {\bibfnamefont {K.~A.}\ \bibnamefont
  {Postnov}},\ }\href {\doibase 10.1103/PhysRevD.90.062008} {\bibfield
  {journal} {\bibinfo  {journal} {Phys. Rev. D}\ }\textbf {\bibinfo {volume}
  {{90}}},\ \bibinfo {pages} {062008} (\bibinfo {year} {{2014}})}\BibitemShut
  {NoStop}%
\bibitem [{\citenamefont {Graham}\ \emph {et~al.}(2016)\citenamefont {Graham},
  \citenamefont {Kaplan}, \citenamefont {Mardon}, \citenamefont {Rajendran},\
  and\ \citenamefont {Terrano}}]{Graham2016}%
  \BibitemOpen
  \bibfield  {author} {\bibinfo {author} {\bibfnamefont {P.~W.}\ \bibnamefont
  {Graham}}, \bibinfo {author} {\bibfnamefont {D.~E.}\ \bibnamefont {Kaplan}},
  \bibinfo {author} {\bibfnamefont {J.}~\bibnamefont {Mardon}}, \bibinfo
  {author} {\bibfnamefont {S.}~\bibnamefont {Rajendran}}, \ and\ \bibinfo
  {author} {\bibfnamefont {W.~A.}\ \bibnamefont {Terrano}},\ }\href {\doibase
  ggf3rq} {\bibfield  {journal} {\bibinfo  {journal} {Phys. Rev. D}\ }\textbf
  {\bibinfo {volume} {{93}}},\ \bibinfo {pages} {075029} (\bibinfo {year}
  {{2016}})}\BibitemShut {NoStop}%
\bibitem [{\citenamefont {de~Martino}\ \emph {et~al.}(2017)\citenamefont
  {de~Martino}, \citenamefont {Broadhurst}, \citenamefont {Tye}, \citenamefont
  {Chiueh}, \citenamefont {Schive},\ and\ \citenamefont
  {Lazkoz}}]{DeMartino2017}%
  \BibitemOpen
  \bibfield  {author} {\bibinfo {author} {\bibfnamefont {I.}~\bibnamefont
  {de~Martino}}, \bibinfo {author} {\bibfnamefont {T.}~\bibnamefont
  {Broadhurst}}, \bibinfo {author} {\bibfnamefont {S.~H.~H.}\ \bibnamefont
  {Tye}}, \bibinfo {author} {\bibfnamefont {T.}~\bibnamefont {Chiueh}},
  \bibinfo {author} {\bibfnamefont {H.-Y.}\ \bibnamefont {Schive}}, \ and\
  \bibinfo {author} {\bibfnamefont {R.}~\bibnamefont {Lazkoz}},\ }\href
  {\doibase 10.1103/PhysRevLett.119.221103} {\bibfield  {journal} {\bibinfo
  {journal} {Phys. Rev. Lett.}\ }\textbf {\bibinfo {volume} {{119}}},\ \bibinfo
  {pages} {221103} (\bibinfo {year} {{2017}})}\BibitemShut {NoStop}%
\bibitem [{\citenamefont {de~Martino}\ \emph {et~al.}(2018)\citenamefont
  {de~Martino}, \citenamefont {Broadhurst}, \citenamefont {Tye}, \citenamefont
  {Chiueh}, \citenamefont {Schive},\ and\ \citenamefont
  {Lazkoz}}]{DeMartino2018}%
  \BibitemOpen
  \bibfield  {author} {\bibinfo {author} {\bibfnamefont {I.}~\bibnamefont
  {de~Martino}}, \bibinfo {author} {\bibfnamefont {T.}~\bibnamefont
  {Broadhurst}}, \bibinfo {author} {\bibfnamefont {S.~H.~H.}\ \bibnamefont
  {Tye}}, \bibinfo {author} {\bibfnamefont {T.}~\bibnamefont {Chiueh}},
  \bibinfo {author} {\bibfnamefont {H.-Y.}\ \bibnamefont {Schive}}, \ and\
  \bibinfo {author} {\bibfnamefont {R.}~\bibnamefont {Lazkoz}},\ }\href
  {\doibase 10.3390/galaxies6010010} {\bibfield  {journal} {\bibinfo  {journal}
  {Galaxies}\ }\textbf {\bibinfo {volume} {{6}}},\ \bibinfo {pages} {10}
  (\bibinfo {year} {{2018}})}\BibitemShut {NoStop}%
\bibitem [{\citenamefont {Cai}\ \emph {et~al.}(2018)\citenamefont {Cai},
  \citenamefont {Liu},\ and\ \citenamefont {Wang}}]{Cai2018}%
  \BibitemOpen
  \bibfield  {author} {\bibinfo {author} {\bibfnamefont {R.-G.}\ \bibnamefont
  {Cai}}, \bibinfo {author} {\bibfnamefont {T.-B.}\ \bibnamefont {Liu}}, \ and\
  \bibinfo {author} {\bibfnamefont {S.-J.}\ \bibnamefont {Wang}},\ }\href
  {\doibase 10.1103/PhysRevD.97.023027} {\bibfield  {journal} {\bibinfo
  {journal} {Phys. Rev. D}\ }\textbf {\bibinfo {volume} {{97}}},\ \bibinfo
  {pages} {023027} (\bibinfo {year} {{2018}})}\BibitemShut {NoStop}%
\bibitem [{\citenamefont {Porayko}\ \emph {et~al.}(2018)\citenamefont
  {Porayko}, \citenamefont {Zhu}, \citenamefont {Levin}, \citenamefont {Hui},
  \citenamefont {Hobbs}, \citenamefont {Grudskaya}, \citenamefont {Postnov},
  \citenamefont {Bailes}, \citenamefont {Bhat}, \citenamefont {Coles} \emph
  {et~al.}}]{Porayko2018}%
  \BibitemOpen
  \bibfield  {author} {\bibinfo {author} {\bibfnamefont {N.~K.}\ \bibnamefont
  {Porayko}}, \bibinfo {author} {\bibfnamefont {X.}~\bibnamefont {Zhu}},
  \bibinfo {author} {\bibfnamefont {Y.}~\bibnamefont {Levin}}, \bibinfo
  {author} {\bibfnamefont {L.}~\bibnamefont {Hui}}, \bibinfo {author}
  {\bibfnamefont {G.}~\bibnamefont {Hobbs}}, \bibinfo {author} {\bibfnamefont
  {A.}~\bibnamefont {Grudskaya}}, \bibinfo {author} {\bibfnamefont
  {K.}~\bibnamefont {Postnov}}, \bibinfo {author} {\bibfnamefont
  {M.}~\bibnamefont {Bailes}}, \bibinfo {author} {\bibfnamefont {N.~D.~R.}\
  \bibnamefont {Bhat}}, \bibinfo {author} {\bibfnamefont {W.}~\bibnamefont
  {Coles}},  \emph {et~al.},\ }\href {\doibase 10.1103/PhysRevD.98.102002}
  {\bibfield  {journal} {\bibinfo  {journal} {Phys. Rev. D}\ }\textbf {\bibinfo
  {volume} {{98}}},\ \bibinfo {pages} {102002} (\bibinfo {year}
  {{2018}})}\BibitemShut {NoStop}%
\bibitem [{\citenamefont {Caputo}\ \emph {et~al.}(2019)\citenamefont {Caputo},
  \citenamefont {Sberna}, \citenamefont {Frias}, \citenamefont {Blas},
  \citenamefont {Pani}, \citenamefont {Shao},\ and\ \citenamefont
  {Yan}}]{Caputo2019}%
  \BibitemOpen
  \bibfield  {author} {\bibinfo {author} {\bibfnamefont {A.}~\bibnamefont
  {Caputo}}, \bibinfo {author} {\bibfnamefont {L.}~\bibnamefont {Sberna}},
  \bibinfo {author} {\bibfnamefont {M.}~\bibnamefont {Frias}}, \bibinfo
  {author} {\bibfnamefont {D.}~\bibnamefont {Blas}}, \bibinfo {author}
  {\bibfnamefont {P.}~\bibnamefont {Pani}}, \bibinfo {author} {\bibfnamefont
  {L.}~\bibnamefont {Shao}}, \ and\ \bibinfo {author} {\bibfnamefont
  {W.}~\bibnamefont {Yan}},\ }\href {\doibase 10.1103/PhysRevD.100.063515}
  {\bibfield  {journal} {\bibinfo  {journal} {Phys. Rev. D}\ }\textbf {\bibinfo
  {volume} {{100}}},\ \bibinfo {pages} {063515} (\bibinfo {year}
  {{2019}})}\BibitemShut {NoStop}%
\bibitem [{\citenamefont {Blas}\ \emph {et~al.}(2020)\citenamefont {Blas},
  \citenamefont {Lopez~Nacir},\ and\ \citenamefont {Sibiryakov}}]{Blas2020}%
  \BibitemOpen
  \bibfield  {author} {\bibinfo {author} {\bibfnamefont {D.}~\bibnamefont
  {Blas}}, \bibinfo {author} {\bibfnamefont {D.}~\bibnamefont {Lopez~Nacir}}, \
  and\ \bibinfo {author} {\bibfnamefont {S.}~\bibnamefont {Sibiryakov}},\
  }\href {\doibase 10.1103/PhysRevD.101.063016} {\bibfield  {journal} {\bibinfo
   {journal} {Phys. Rev. D}\ }\textbf {\bibinfo {volume} {{101}}},\ \bibinfo
  {pages} {063016} (\bibinfo {year} {{2020}})}\BibitemShut {NoStop}%
\bibitem [{\citenamefont {Siegel}\ \emph {et~al.}(2007)\citenamefont {Siegel},
  \citenamefont {Hertzberg},\ and\ \citenamefont {Fry}}]{Siegel2007}%
  \BibitemOpen
  \bibfield  {author} {\bibinfo {author} {\bibfnamefont {E.~R.}\ \bibnamefont
  {Siegel}}, \bibinfo {author} {\bibfnamefont {M.~P.}\ \bibnamefont
  {Hertzberg}}, \ and\ \bibinfo {author} {\bibfnamefont {J.~N.}\ \bibnamefont
  {Fry}},\ }\href {\doibase 10.1111/j.1365-2966.2007.12435.x} {\bibfield
  {journal} {\bibinfo  {journal} {Mon. Not. R. Astron. Soc.}\ }\textbf
  {\bibinfo {volume} {382}},\ \bibinfo {pages} {879} (\bibinfo {year}
  {2007})}\BibitemShut {NoStop}%
\bibitem [{\citenamefont {Baghram}\ \emph {et~al.}(2011)\citenamefont
  {Baghram}, \citenamefont {Afshordi},\ and\ \citenamefont
  {Zurek}}]{Baghram2011}%
  \BibitemOpen
  \bibfield  {author} {\bibinfo {author} {\bibfnamefont {S.}~\bibnamefont
  {Baghram}}, \bibinfo {author} {\bibfnamefont {N.}~\bibnamefont {Afshordi}}, \
  and\ \bibinfo {author} {\bibfnamefont {K.~M.}\ \bibnamefont {Zurek}},\ }\href
  {\doibase 10.1103/PhysRevD.84.043511} {\bibfield  {journal} {\bibinfo
  {journal} {Phys. Rev. D}\ }\textbf {\bibinfo {volume} {84}},\ \bibinfo
  {pages} {043511} (\bibinfo {year} {2011})}\BibitemShut {NoStop}%
\bibitem [{\citenamefont {Clark}\ \emph
  {et~al.}(2015{\natexlab{a}})\citenamefont {Clark}, \citenamefont {Lewis},\
  and\ \citenamefont {Scott}}]{Clark2015}%
  \BibitemOpen
  \bibfield  {author} {\bibinfo {author} {\bibfnamefont {H.~A.}\ \bibnamefont
  {Clark}}, \bibinfo {author} {\bibfnamefont {G.~F.}\ \bibnamefont {Lewis}}, \
  and\ \bibinfo {author} {\bibfnamefont {P.}~\bibnamefont {Scott}},\ }\href
  {\doibase 10.1093/mnras/stv2743} {\bibfield  {journal} {\bibinfo  {journal}
  {Mon. Not. R. Astron. Soc.}\ }\textbf {\bibinfo {volume} {456}},\ \bibinfo
  {pages} {1394} (\bibinfo {year} {2015}{\natexlab{a}})}\BibitemShut {NoStop}%
\bibitem [{\citenamefont {Clark}\ \emph
  {et~al.}(2015{\natexlab{b}})\citenamefont {Clark}, \citenamefont {Lewis},\
  and\ \citenamefont {Scott}}]{Clark2015a}%
  \BibitemOpen
  \bibfield  {author} {\bibinfo {author} {\bibfnamefont {H.~A.}\ \bibnamefont
  {Clark}}, \bibinfo {author} {\bibfnamefont {G.~F.}\ \bibnamefont {Lewis}}, \
  and\ \bibinfo {author} {\bibfnamefont {P.}~\bibnamefont {Scott}},\ }\href
  {\doibase 10.1093/mnras/stv2529} {\bibfield  {journal} {\bibinfo  {journal}
  {Mon. Not. R. Astron. Soc.}\ }\textbf {\bibinfo {volume} {456}},\ \bibinfo
  {pages} {1402} (\bibinfo {year} {2015}{\natexlab{b}})}\BibitemShut {NoStop}%
\bibitem [{\citenamefont {Ramani}\ \emph {et~al.}(2020)\citenamefont {Ramani},
  \citenamefont {Trickle},\ and\ \citenamefont {Zurek}}]{Ramani2020}%
  \BibitemOpen
  \bibfield  {author} {\bibinfo {author} {\bibfnamefont {H.}~\bibnamefont
  {Ramani}}, \bibinfo {author} {\bibfnamefont {T.}~\bibnamefont {Trickle}}, \
  and\ \bibinfo {author} {\bibfnamefont {K.~M.}\ \bibnamefont {Zurek}},\
  }\href@noop {} {} (\bibinfo {year} {2020}),\ \Eprint
  {http://arxiv.org/abs/2005.03030} {arXiv:2005.03030 [astro-ph.CO]}
  \BibitemShut {NoStop}%
\bibitem [{\citenamefont {Damour}\ and\ \citenamefont
  {Taylor}(1991)}]{Damour1991}%
  \BibitemOpen
  \bibfield  {author} {\bibinfo {author} {\bibfnamefont {T.}~\bibnamefont
  {Damour}}\ and\ \bibinfo {author} {\bibfnamefont {J.~H.}\ \bibnamefont
  {Taylor}},\ }\href {\doibase 10.1086/169585} {\bibfield  {journal} {\bibinfo
  {journal} {Astrophys. J.}\ }\textbf {\bibinfo {volume} {366}},\ \bibinfo
  {pages} {501} (\bibinfo {year} {1991})}\BibitemShut {NoStop}%
\bibitem [{\citenamefont {Phinney}\ \emph {et~al.}(1992)\citenamefont
  {Phinney}, \citenamefont {Blandford}, \citenamefont {Hewish}, \citenamefont
  {Lyne},\ and\ \citenamefont {Mestel}}]{Phinney1992}%
  \BibitemOpen
  \bibfield  {author} {\bibinfo {author} {\bibfnamefont {E.~S.}\ \bibnamefont
  {Phinney}}, \bibinfo {author} {\bibfnamefont {R.~D.}\ \bibnamefont
  {Blandford}}, \bibinfo {author} {\bibfnamefont {A.}~\bibnamefont {Hewish}},
  \bibinfo {author} {\bibfnamefont {A.~G.}\ \bibnamefont {Lyne}}, \ and\
  \bibinfo {author} {\bibfnamefont {L.}~\bibnamefont {Mestel}},\ }\href
  {\doibase 10.1098/rsta.1992.0084} {\bibfield  {journal} {\bibinfo  {journal}
  {Philos. Trans. R. Soc. A}\ }\textbf {\bibinfo {volume} {341}},\ \bibinfo
  {pages} {39} (\bibinfo {year} {1992})}\BibitemShut {NoStop}%
\bibitem [{\citenamefont {Nice}\ and\ \citenamefont {Taylor}(1995)}]{Nice1995}%
  \BibitemOpen
  \bibfield  {author} {\bibinfo {author} {\bibfnamefont {D.~J.}\ \bibnamefont
  {Nice}}\ and\ \bibinfo {author} {\bibfnamefont {J.~H.}\ \bibnamefont
  {Taylor}},\ }\href {\doibase 10.1086/175367} {\bibfield  {journal} {\bibinfo
  {journal} {Astrophys. J.}\ }\textbf {\bibinfo {volume} {441}},\ \bibinfo
  {pages} {429} (\bibinfo {year} {1995})}\BibitemShut {NoStop}%
\bibitem [{\citenamefont {{Shklovskii}}(1970)}]{Shklovskii1970}%
  \BibitemOpen
  \bibfield  {author} {\bibinfo {author} {\bibfnamefont {I.~S.}\ \bibnamefont
  {{Shklovskii}}},\ }\href@noop {} {\bibfield  {journal} {\bibinfo  {journal}
  {Soviet~Ast.}\ }\textbf {\bibinfo {volume} {13}},\ \bibinfo {pages} {562}
  (\bibinfo {year} {1970})}\BibitemShut {NoStop}%
\bibitem [{\citenamefont {Lyne}\ and\ \citenamefont
  {Graham-Smith}(2012)}]{Lyne2012}%
  \BibitemOpen
  \bibfield  {author} {\bibinfo {author} {\bibfnamefont {A.}~\bibnamefont
  {Lyne}}\ and\ \bibinfo {author} {\bibfnamefont {F.}~\bibnamefont
  {Graham-Smith}},\ }\href@noop {} {\emph {\bibinfo {title} {Pulsar
  Astronomy}}},\ \bibinfo {edition} {4th}\ ed.\ (\bibinfo  {publisher}
  {Cambridge University Press},\ \bibinfo {year} {2012})\BibitemShut {NoStop}%
\bibitem [{\citenamefont {Xu}\ and\ \citenamefont {Qiao}(2001)}]{Xu2001}%
  \BibitemOpen
  \bibfield  {author} {\bibinfo {author} {\bibfnamefont {R.~X.}\ \bibnamefont
  {Xu}}\ and\ \bibinfo {author} {\bibfnamefont {G.~J.}\ \bibnamefont {Qiao}},\
  }\href {\doibase 10.1086/324381} {\bibfield  {journal} {\bibinfo  {journal}
  {Astrophys. J.}\ }\textbf {\bibinfo {volume} {561}},\ \bibinfo {pages} {L85}
  (\bibinfo {year} {2001})}\BibitemShut {NoStop}%
\bibitem [{\citenamefont {Peters}\ and\ \citenamefont
  {Mathews}(1963)}]{Peters:1963}%
  \BibitemOpen
  \bibfield  {author} {\bibinfo {author} {\bibfnamefont {P.~C.}\ \bibnamefont
  {Peters}}\ and\ \bibinfo {author} {\bibfnamefont {J.}~\bibnamefont
  {Mathews}},\ }\href {\doibase 10.1103/PhysRev.131.435} {\bibfield  {journal}
  {\bibinfo  {journal} {Phys. Rev.}\ }\textbf {\bibinfo {volume} {131}},\
  \bibinfo {pages} {435} (\bibinfo {year} {1963})}\BibitemShut {NoStop}%
\bibitem [{SM()}]{SM}%
  \BibitemOpen
  \href@noop {} {\ }\bibinfo {note} {See Supplemental Material {\tt [URL]} for
  details on spin and orbital period analyses. The SM includes references
  \cite{Lyne2012,YMW16,Hogg2010,DFM2013,Shao:2020fka,Peters:1963,Antoniadis:2013pzd,Reardon:2015kba,Perera:2019sca,Desvignes:2016yex,lazaridis2009,Fonseca:2014qla,Zhu:2018etc,Freire:2012mg,Ferdman:2014rna,Weisberg2016,Weisberg2008,Cognard:2017xyr,Deller2018,Haniewicz:2020jro}}\BibitemShut
  {NoStop}%
\bibitem [{\citenamefont {Sofue}\ \emph {et~al.}(2009)\citenamefont {Sofue},
  \citenamefont {Honma},\ and\ \citenamefont {Omodaka}}]{Sofue2009}%
  \BibitemOpen
  \bibfield  {author} {\bibinfo {author} {\bibfnamefont {Y.}~\bibnamefont
  {Sofue}}, \bibinfo {author} {\bibfnamefont {M.}~\bibnamefont {Honma}}, \ and\
  \bibinfo {author} {\bibfnamefont {T.}~\bibnamefont {Omodaka}},\ }\href
  {\doibase 10.1093/pasj/61.2.227} {\bibfield  {journal} {\bibinfo  {journal}
  {Publ. Astron. Soc. Jpn.}\ }\textbf {\bibinfo {volume} {61}},\ \bibinfo
  {pages} {227} (\bibinfo {year} {2009})},\ \bibinfo {note} {{Complete data
  tables available
  \href{http://www.ioa.s.u-tokyo.ac.jp/~sofue/mw/rc2009/}{online}}}\BibitemShut
  {NoStop}%
\bibitem [{\citenamefont {Read}(2014)}]{Read2014}%
  \BibitemOpen
  \bibfield  {author} {\bibinfo {author} {\bibfnamefont {J.~I.}\ \bibnamefont
  {Read}},\ }\href {\doibase 10.1088/0954-3899/41/6/063101} {\bibfield
  {journal} {\bibinfo  {journal} {J. Phys. G: Nucl. Part. Phys.}\ }\textbf
  {\bibinfo {volume} {41}},\ \bibinfo {pages} {063101} (\bibinfo {year}
  {2014})}\BibitemShut {NoStop}%
\bibitem [{\citenamefont {Manchester}\ \emph {et~al.}(2005)\citenamefont
  {Manchester}, \citenamefont {Hobbs}, \citenamefont {Teoh},\ and\
  \citenamefont {Hobbs}}]{Manchester2005}%
  \BibitemOpen
  \bibfield  {author} {\bibinfo {author} {\bibfnamefont {R.~N.}\ \bibnamefont
  {Manchester}}, \bibinfo {author} {\bibfnamefont {G.~B.}\ \bibnamefont
  {Hobbs}}, \bibinfo {author} {\bibfnamefont {A.}~\bibnamefont {Teoh}}, \ and\
  \bibinfo {author} {\bibfnamefont {M.}~\bibnamefont {Hobbs}},\ }\href
  {\doibase 10.1086/428488} {\bibfield  {journal} {\bibinfo  {journal} {Astron.
  J.}\ }\textbf {\bibinfo {volume} {129}},\ \bibinfo {pages} {1993} (\bibinfo
  {year} {2005})}\BibitemShut {NoStop}%
\bibitem [{\citenamefont {Evans}\ \emph {et~al.}(2019)\citenamefont {Evans},
  \citenamefont {O'Hare},\ and\ \citenamefont {McCabe}}]{Evans:2018bqy}%
  \BibitemOpen
  \bibfield  {author} {\bibinfo {author} {\bibfnamefont {N.~W.}\ \bibnamefont
  {Evans}}, \bibinfo {author} {\bibfnamefont {C.~A.}\ \bibnamefont {O'Hare}}, \
  and\ \bibinfo {author} {\bibfnamefont {C.}~\bibnamefont {McCabe}},\ }\href
  {\doibase 10.1103/PhysRevD.99.023012} {\bibfield  {journal} {\bibinfo
  {journal} {Phys. Rev. D}\ }\textbf {\bibinfo {volume} {99}},\ \bibinfo
  {pages} {023012} (\bibinfo {year} {2019})}\BibitemShut {NoStop}%
\bibitem [{\citenamefont {McMillan}(2016)}]{McMillan2016}%
  \BibitemOpen
  \bibfield  {author} {\bibinfo {author} {\bibfnamefont {P.~J.}\ \bibnamefont
  {McMillan}},\ }\href {\doibase 10.1093/mnras/stw2759} {\bibfield  {journal}
  {\bibinfo  {journal} {Mon. Not. R. Astron. Soc.}\ }\textbf {\bibinfo {volume}
  {465}},\ \bibinfo {pages} {76} (\bibinfo {year} {2016})}\BibitemShut
  {NoStop}%
\bibitem [{\citenamefont {Abuter}\ \emph {et~al.}(2019)\citenamefont {Abuter}
  \emph {et~al.}}]{abuter_geometric_2019}%
  \BibitemOpen
  \bibfield  {author} {\bibinfo {author} {\bibfnamefont {R.}~\bibnamefont
  {Abuter}} \emph {et~al.} (\bibinfo {collaboration} {The GRAVITY
  Collaboration}),\ }\href {\doibase 10.1051/0004-6361/201935656} {\bibfield
  {journal} {\bibinfo  {journal} {Astron. Astrophys.}\ }\textbf {\bibinfo
  {volume} {625}},\ \bibinfo {pages} {L10} (\bibinfo {year}
  {2019})}\BibitemShut {NoStop}%
\bibitem [{\citenamefont {Bates}\ \emph {et~al.}(2011)\citenamefont {Bates},
  \citenamefont {Bailes}, \citenamefont {Bhat}, \citenamefont {Burgay},
  \citenamefont {Burke-Spolaor}, \citenamefont {D'Amico}, \citenamefont
  {Jameson}, \citenamefont {Johnston}, \citenamefont {Keith}, \citenamefont
  {Kramer} \emph {et~al.}}]{Bates2011}%
  \BibitemOpen
  \bibfield  {author} {\bibinfo {author} {\bibfnamefont {S.~D.}\ \bibnamefont
  {Bates}}, \bibinfo {author} {\bibfnamefont {M.}~\bibnamefont {Bailes}},
  \bibinfo {author} {\bibfnamefont {N.~D.~R.}\ \bibnamefont {Bhat}}, \bibinfo
  {author} {\bibfnamefont {M.}~\bibnamefont {Burgay}}, \bibinfo {author}
  {\bibfnamefont {S.}~\bibnamefont {Burke-Spolaor}}, \bibinfo {author}
  {\bibfnamefont {N.}~\bibnamefont {D'Amico}}, \bibinfo {author} {\bibfnamefont
  {A.}~\bibnamefont {Jameson}}, \bibinfo {author} {\bibfnamefont
  {S.}~\bibnamefont {Johnston}}, \bibinfo {author} {\bibfnamefont {M.~J.}\
  \bibnamefont {Keith}}, \bibinfo {author} {\bibfnamefont {M.}~\bibnamefont
  {Kramer}},  \emph {et~al.},\ }\href {\doibase 10/bffd5w} {\bibfield
  {journal} {\bibinfo  {journal} {Mon. Not. R. Astron. Soc.}\ }\textbf
  {\bibinfo {volume} {416}},\ \bibinfo {pages} {2455} (\bibinfo {year}
  {2011})}\BibitemShut {NoStop}%
\bibitem [{\citenamefont {Kuijken}\ and\ \citenamefont
  {Gilmore}(1989)}]{Kuijken1989}%
  \BibitemOpen
  \bibfield  {author} {\bibinfo {author} {\bibfnamefont {K.}~\bibnamefont
  {Kuijken}}\ and\ \bibinfo {author} {\bibfnamefont {G.}~\bibnamefont
  {Gilmore}},\ }\href {\doibase 10.1093/mnras/239.2.605} {\bibfield  {journal}
  {\bibinfo  {journal} {Mon. Not. R. Astron. Soc.}\ }\textbf {\bibinfo {volume}
  {239}},\ \bibinfo {pages} {605} (\bibinfo {year} {1989})}\BibitemShut
  {NoStop}%
\bibitem [{\citenamefont {Shao}\ \emph {et~al.}(2020)\citenamefont {Shao},
  \citenamefont {Wex},\ and\ \citenamefont {Zhou}}]{Shao:2020fka}%
  \BibitemOpen
  \bibfield  {author} {\bibinfo {author} {\bibfnamefont {L.}~\bibnamefont
  {Shao}}, \bibinfo {author} {\bibfnamefont {N.}~\bibnamefont {Wex}}, \ and\
  \bibinfo {author} {\bibfnamefont {S.-Y.}\ \bibnamefont {Zhou}},\ }\href
  {\doibase 10.1103/PhysRevD.102.024069} {\bibfield  {journal} {\bibinfo
  {journal} {Phys. Rev. D}\ }\textbf {\bibinfo {volume} {102}},\ \bibinfo
  {pages} {024069} (\bibinfo {year} {2020})}\BibitemShut {NoStop}%
\bibitem [{\citenamefont {McKee}\ \emph {et~al.}(2015)\citenamefont {McKee},
  \citenamefont {Parravano},\ and\ \citenamefont {Hollenbach}}]{Mckee2015}%
  \BibitemOpen
  \bibfield  {author} {\bibinfo {author} {\bibfnamefont {C.~F.}\ \bibnamefont
  {McKee}}, \bibinfo {author} {\bibfnamefont {A.}~\bibnamefont {Parravano}}, \
  and\ \bibinfo {author} {\bibfnamefont {D.~J.}\ \bibnamefont {Hollenbach}},\
  }\href {\doibase 10.1088/0004-637X/814/1/13} {\bibfield  {journal} {\bibinfo
  {journal} {Astrophys. J. Lett.}\ }\textbf {\bibinfo {volume} {814}},\
  \bibinfo {pages} {13} (\bibinfo {year} {2015})}\BibitemShut {NoStop}%
\bibitem [{\citenamefont {Buckley}\ and\ \citenamefont
  {Peter}(2018)}]{Buckley:2018}%
  \BibitemOpen
  \bibfield  {author} {\bibinfo {author} {\bibfnamefont {M.~R.}\ \bibnamefont
  {Buckley}}\ and\ \bibinfo {author} {\bibfnamefont {A.~H.~G.}\ \bibnamefont
  {Peter}},\ }\href {\doibase https://doi.org/10.1016/j.physrep.2018.07.003}
  {\bibfield  {journal} {\bibinfo  {journal} {Physics Reports}\ }\textbf
  {\bibinfo {volume} {761}},\ \bibinfo {pages} {1 } (\bibinfo {year}
  {2018})}\BibitemShut {NoStop}%
\bibitem [{\citenamefont {Jennings}\ \emph {et~al.}(2018)\citenamefont
  {Jennings}, \citenamefont {Kaplan}, \citenamefont {Chatterjee}, \citenamefont
  {Cordes},\ and\ \citenamefont {Deller}}]{Jennings2018}%
  \BibitemOpen
  \bibfield  {author} {\bibinfo {author} {\bibfnamefont {R.~J.}\ \bibnamefont
  {Jennings}}, \bibinfo {author} {\bibfnamefont {D.~L.}\ \bibnamefont
  {Kaplan}}, \bibinfo {author} {\bibfnamefont {S.}~\bibnamefont {Chatterjee}},
  \bibinfo {author} {\bibfnamefont {J.~M.}\ \bibnamefont {Cordes}}, \ and\
  \bibinfo {author} {\bibfnamefont {A.~T.}\ \bibnamefont {Deller}},\ }\href
  {https://arxiv.org/pdf/1806.06076} {\bibfield  {journal} {\bibinfo  {journal}
  {Astrophys. J.}\ }\textbf {\bibinfo {volume} {864}},\ \bibinfo {pages} {26}
  (\bibinfo {year} {2018})}\BibitemShut {NoStop}%
\bibitem [{\citenamefont {Mingarelli}\ \emph {et~al.}(2018)\citenamefont
  {Mingarelli}, \citenamefont {Anderson}, \citenamefont {Bedell},\ and\
  \citenamefont {Spergel}}]{Mingarelli2018}%
  \BibitemOpen
  \bibfield  {author} {\bibinfo {author} {\bibfnamefont {C.~M.~F.}\
  \bibnamefont {Mingarelli}}, \bibinfo {author} {\bibfnamefont
  {L.}~\bibnamefont {Anderson}}, \bibinfo {author} {\bibfnamefont
  {M.}~\bibnamefont {Bedell}}, \ and\ \bibinfo {author} {\bibfnamefont {D.~N.}\
  \bibnamefont {Spergel}},\ }\href@noop {} {} (\bibinfo {year} {2018}),\
  \Eprint {http://arxiv.org/abs/1812.06262} {arXiv:1812.06262 [astro-ph.IM]}
  \BibitemShut {NoStop}%
\bibitem [{1()}]{1}%
  \BibitemOpen
  \href@noop {} {}\bibinfo {note} {\href {
  https://github.com/R3zaEbadi/MW-Accelerometry}{\tt
  https://github.com/R3zaEbadi/MW-Accelerometry}}\BibitemShut {NoStop}%
\bibitem [{\citenamefont {{Pitkin}}(2018)}]{psrqpy}%
  \BibitemOpen
  \bibfield  {author} {\bibinfo {author} {\bibfnamefont {M.}~\bibnamefont
  {{Pitkin}}},\ }\href {\doibase 10.21105/joss.00538} {\bibfield  {journal}
  {\bibinfo  {journal} {{J. Open Source Softw.}}\ }\textbf {\bibinfo {volume}
  {3}},\ \bibinfo {pages} {538} (\bibinfo {year} {2018})}\BibitemShut {NoStop}%
\bibitem [{\citenamefont {Foreman-Mackey}\ \emph {et~al.}(2013)\citenamefont
  {Foreman-Mackey}, \citenamefont {Hogg}, \citenamefont {Lang},\ and\
  \citenamefont {Goodman}}]{DFM2013}%
  \BibitemOpen
  \bibfield  {author} {\bibinfo {author} {\bibfnamefont {D.}~\bibnamefont
  {Foreman-Mackey}}, \bibinfo {author} {\bibfnamefont {D.~W.}\ \bibnamefont
  {Hogg}}, \bibinfo {author} {\bibfnamefont {D.}~\bibnamefont {Lang}}, \ and\
  \bibinfo {author} {\bibfnamefont {J.}~\bibnamefont {Goodman}},\ }\href
  {\doibase 10.1086/670067} {\bibfield  {journal} {\bibinfo  {journal} {Publ.
  Astron. Soc. Pac.}\ }\textbf {\bibinfo {volume} {125}},\ \bibinfo {pages}
  {306} (\bibinfo {year} {2013})}\BibitemShut {NoStop}%
\bibitem [{\citenamefont {Foreman-Mackey}(2016)}]{corner}%
  \BibitemOpen
  \bibfield  {author} {\bibinfo {author} {\bibfnamefont {D.}~\bibnamefont
  {Foreman-Mackey}},\ }\href {\doibase 10.21105/joss.00024} {\bibfield
  {journal} {\bibinfo  {journal} {J. Open Source Softw.}\ }\textbf {\bibinfo
  {volume} {1}},\ \bibinfo {pages} {24} (\bibinfo {year} {2016})}\BibitemShut
  {NoStop}%
\bibitem [{\citenamefont {Chakrabarti}\ \emph {et~al.}(2021)\citenamefont
  {Chakrabarti}, \citenamefont {Chang}, \citenamefont {Lam}, \citenamefont
  {Vigeland},\ and\ \citenamefont {Quillen}}]{Chakrabarti2021}%
  \BibitemOpen
  \bibfield  {author} {\bibinfo {author} {\bibfnamefont {S.}~\bibnamefont
  {Chakrabarti}}, \bibinfo {author} {\bibfnamefont {P.}~\bibnamefont {Chang}},
  \bibinfo {author} {\bibfnamefont {M.~T.}\ \bibnamefont {Lam}}, \bibinfo
  {author} {\bibfnamefont {S.~J.}\ \bibnamefont {Vigeland}}, \ and\ \bibinfo
  {author} {\bibfnamefont {A.~C.}\ \bibnamefont {Quillen}},\ }\href {\doibase
  10.3847/2041-8213/abd635} {\bibfield  {journal} {\bibinfo  {journal}
  {Astrophys. J. Lett.}\ }\textbf {\bibinfo {volume} {907}},\ \bibinfo {pages}
  {L26} (\bibinfo {year} {2021})}\BibitemShut {NoStop}%
\bibitem [{\citenamefont {Klioner}\ \emph {et~al.}(2020)\citenamefont {Klioner}
  \emph {et~al.}}]{Klioner2020}%
  \BibitemOpen
  \bibfield  {author} {\bibinfo {author} {\bibfnamefont {S.~A.}\ \bibnamefont
  {Klioner}} \emph {et~al.} (\bibinfo {collaboration} {\emph{Gaia}
  Collaboration}),\ }\href@noop {} {\  (\bibinfo {year} {2020})},\ \Eprint
  {http://arxiv.org/abs/2012.02036} {arXiv:2012.02036 [astro-ph.GA]}
  \BibitemShut {NoStop}%
\bibitem [{\citenamefont {Bovy}(2020)}]{Bovy2020}%
  \BibitemOpen
  \bibfield  {author} {\bibinfo {author} {\bibfnamefont {J.}~\bibnamefont
  {Bovy}},\ }\href@noop {} {\  (\bibinfo {year} {2020})},\ \Eprint
  {http://arxiv.org/abs/2012.02169} {arXiv:2012.02169 [astro-ph.GA]}
  \BibitemShut {NoStop}%
\bibitem [{\citenamefont {Yao}\ \emph {et~al.}(2017)\citenamefont {Yao},
  \citenamefont {Manchester},\ and\ \citenamefont {Wang}}]{YMW16}%
  \BibitemOpen
  \bibfield  {author} {\bibinfo {author} {\bibfnamefont {J.~M.}\ \bibnamefont
  {Yao}}, \bibinfo {author} {\bibfnamefont {R.~N.}\ \bibnamefont {Manchester}},
  \ and\ \bibinfo {author} {\bibfnamefont {N.}~\bibnamefont {Wang}},\ }\href
  {\doibase 10.3847/1538-4357/835/1/29} {\bibfield  {journal} {\bibinfo
  {journal} {Astrophys. J.}\ }\textbf {\bibinfo {volume} {835}},\ \bibinfo
  {pages} {29} (\bibinfo {year} {2017})}\BibitemShut {NoStop}%
\bibitem [{\citenamefont {Hogg}\ \emph {et~al.}(2010)\citenamefont {Hogg},
  \citenamefont {Bovy},\ and\ \citenamefont {Lang}}]{Hogg2010}%
  \BibitemOpen
  \bibfield  {author} {\bibinfo {author} {\bibfnamefont {D.~W.}\ \bibnamefont
  {Hogg}}, \bibinfo {author} {\bibfnamefont {J.}~\bibnamefont {Bovy}}, \ and\
  \bibinfo {author} {\bibfnamefont {D.}~\bibnamefont {Lang}},\ }\href@noop {}
  {} (\bibinfo {year} {2010}),\ \Eprint {http://arxiv.org/abs/1008.4686}
  {arXiv:1008.4686 [astro-ph.IM]} \BibitemShut {NoStop}%
\bibitem [{\citenamefont {Antoniadis}\ \emph {et~al.}(2013)\citenamefont
  {Antoniadis}, \citenamefont {Freire}, \citenamefont {Wex}, \citenamefont
  {Tauris}, \citenamefont {Lynch}, \citenamefont {van Kerkwijk}, \citenamefont
  {Kramer}, \citenamefont {Bassa}, \citenamefont {Dhillon}, \citenamefont
  {Driebe} \emph {et~al.}}]{Antoniadis:2013pzd}%
  \BibitemOpen
  \bibfield  {author} {\bibinfo {author} {\bibfnamefont {J.}~\bibnamefont
  {Antoniadis}}, \bibinfo {author} {\bibfnamefont {P.~C.}\ \bibnamefont
  {Freire}}, \bibinfo {author} {\bibfnamefont {N.}~\bibnamefont {Wex}},
  \bibinfo {author} {\bibfnamefont {T.~M.}\ \bibnamefont {Tauris}}, \bibinfo
  {author} {\bibfnamefont {R.~S.}\ \bibnamefont {Lynch}}, \bibinfo {author}
  {\bibfnamefont {M.~H.}\ \bibnamefont {van Kerkwijk}}, \bibinfo {author}
  {\bibfnamefont {M.}~\bibnamefont {Kramer}}, \bibinfo {author} {\bibfnamefont
  {C.}~\bibnamefont {Bassa}}, \bibinfo {author} {\bibfnamefont {V.~S.}\
  \bibnamefont {Dhillon}}, \bibinfo {author} {\bibfnamefont {T.}~\bibnamefont
  {Driebe}},  \emph {et~al.},\ }\href {\doibase 10.1126/science.1233232}
  {\bibfield  {journal} {\bibinfo  {journal} {Science}\ }\textbf {\bibinfo
  {volume} {340}},\ \bibinfo {pages} {6131} (\bibinfo {year}
  {2013})}\BibitemShut {NoStop}%
\bibitem [{\citenamefont {Reardon}\ \emph {et~al.}(2016)\citenamefont
  {Reardon}, \citenamefont {Hobbs}, \citenamefont {Coles}, \citenamefont
  {Levin}, \citenamefont {Keith}, \citenamefont {Bailes}, \citenamefont {Bhat},
  \citenamefont {Burke-Spolaor}, \citenamefont {Dai}, \citenamefont {Kerr}
  \emph {et~al.}}]{Reardon:2015kba}%
  \BibitemOpen
  \bibfield  {author} {\bibinfo {author} {\bibfnamefont {D.}~\bibnamefont
  {Reardon}}, \bibinfo {author} {\bibfnamefont {G.}~\bibnamefont {Hobbs}},
  \bibinfo {author} {\bibfnamefont {W.}~\bibnamefont {Coles}}, \bibinfo
  {author} {\bibfnamefont {Y.}~\bibnamefont {Levin}}, \bibinfo {author}
  {\bibfnamefont {M.}~\bibnamefont {Keith}}, \bibinfo {author} {\bibfnamefont
  {M.}~\bibnamefont {Bailes}}, \bibinfo {author} {\bibfnamefont
  {N.}~\bibnamefont {Bhat}}, \bibinfo {author} {\bibfnamefont {S.}~\bibnamefont
  {Burke-Spolaor}}, \bibinfo {author} {\bibfnamefont {S.}~\bibnamefont {Dai}},
  \bibinfo {author} {\bibfnamefont {M.}~\bibnamefont {Kerr}},  \emph {et~al.},\
  }\href {\doibase 10.1093/mnras/stv2395} {\bibfield  {journal} {\bibinfo
  {journal} {Mon. Not. Roy. Astron. Soc.}\ }\textbf {\bibinfo {volume} {455}},\
  \bibinfo {pages} {1751} (\bibinfo {year} {2016})}\BibitemShut {NoStop}%
\bibitem [{\citenamefont {Perera}\ \emph {et~al.}(2019)\citenamefont {Perera},
  \citenamefont {DeCesar}, \citenamefont {Demorest}, \citenamefont {Kerr},
  \citenamefont {Lentati}, \citenamefont {Nice}, \citenamefont {Os{\l}owski},
  \citenamefont {Ransom}, \citenamefont {Keith}, \citenamefont {Arzoumanian}
  \emph {et~al.}}]{Perera:2019sca}%
  \BibitemOpen
  \bibfield  {author} {\bibinfo {author} {\bibfnamefont {B.}~\bibnamefont
  {Perera}}, \bibinfo {author} {\bibfnamefont {M.}~\bibnamefont {DeCesar}},
  \bibinfo {author} {\bibfnamefont {P.}~\bibnamefont {Demorest}}, \bibinfo
  {author} {\bibfnamefont {M.}~\bibnamefont {Kerr}}, \bibinfo {author}
  {\bibfnamefont {L.}~\bibnamefont {Lentati}}, \bibinfo {author} {\bibfnamefont
  {D.}~\bibnamefont {Nice}}, \bibinfo {author} {\bibfnamefont {S.}~\bibnamefont
  {Os{\l}owski}}, \bibinfo {author} {\bibfnamefont {S.}~\bibnamefont {Ransom}},
  \bibinfo {author} {\bibfnamefont {M.}~\bibnamefont {Keith}}, \bibinfo
  {author} {\bibfnamefont {Z.}~\bibnamefont {Arzoumanian}},  \emph {et~al.},\
  }\href {\doibase 10.1093/mnras/stz2857} {\bibfield  {journal} {\bibinfo
  {journal} {Mon. Not. Roy. Astron. Soc.}\ }\textbf {\bibinfo {volume} {490}},\
  \bibinfo {pages} {4666} (\bibinfo {year} {2019})}\BibitemShut {NoStop}%
\bibitem [{\citenamefont {Desvignes}\ \emph {et~al.}(2016)\citenamefont
  {Desvignes}, \citenamefont {Caballero}, \citenamefont {Lentati},
  \citenamefont {Verbiest}, \citenamefont {Champion}, \citenamefont {Stappers},
  \citenamefont {Janssen}, \citenamefont {Lazarus}, \citenamefont
  {Os{\l}owski}, \citenamefont {Babak} \emph {et~al.}}]{Desvignes:2016yex}%
  \BibitemOpen
  \bibfield  {author} {\bibinfo {author} {\bibfnamefont {G.}~\bibnamefont
  {Desvignes}}, \bibinfo {author} {\bibfnamefont {R.}~\bibnamefont
  {Caballero}}, \bibinfo {author} {\bibfnamefont {L.}~\bibnamefont {Lentati}},
  \bibinfo {author} {\bibfnamefont {J.}~\bibnamefont {Verbiest}}, \bibinfo
  {author} {\bibfnamefont {D.}~\bibnamefont {Champion}}, \bibinfo {author}
  {\bibfnamefont {B.}~\bibnamefont {Stappers}}, \bibinfo {author}
  {\bibfnamefont {G.}~\bibnamefont {Janssen}}, \bibinfo {author} {\bibfnamefont
  {P.}~\bibnamefont {Lazarus}}, \bibinfo {author} {\bibfnamefont
  {S.}~\bibnamefont {Os{\l}owski}}, \bibinfo {author} {\bibfnamefont
  {S.}~\bibnamefont {Babak}},  \emph {et~al.},\ }\href {\doibase 10/f8n4n8}
  {\bibfield  {journal} {\bibinfo  {journal} {Mon. Not. Roy. Astron. Soc.}\
  }\textbf {\bibinfo {volume} {458}},\ \bibinfo {pages} {3341} (\bibinfo {year}
  {2016})}\BibitemShut {NoStop}%
\bibitem [{\citenamefont {Lazaridis}\ \emph {et~al.}(2009)\citenamefont
  {Lazaridis}, \citenamefont {Wex}, \citenamefont {Jessner}, \citenamefont
  {Kramer}, \citenamefont {Stappers}, \citenamefont {Janssen}, \citenamefont
  {Desvignes}, \citenamefont {Purver}, \citenamefont {Cognard}, \citenamefont
  {Theureau} \emph {et~al.}}]{lazaridis2009}%
  \BibitemOpen
  \bibfield  {author} {\bibinfo {author} {\bibfnamefont {K.}~\bibnamefont
  {Lazaridis}}, \bibinfo {author} {\bibfnamefont {N.}~\bibnamefont {Wex}},
  \bibinfo {author} {\bibfnamefont {A.}~\bibnamefont {Jessner}}, \bibinfo
  {author} {\bibfnamefont {M.}~\bibnamefont {Kramer}}, \bibinfo {author}
  {\bibfnamefont {B.}~\bibnamefont {Stappers}}, \bibinfo {author}
  {\bibfnamefont {G.}~\bibnamefont {Janssen}}, \bibinfo {author} {\bibfnamefont
  {G.}~\bibnamefont {Desvignes}}, \bibinfo {author} {\bibfnamefont
  {M.}~\bibnamefont {Purver}}, \bibinfo {author} {\bibfnamefont
  {I.}~\bibnamefont {Cognard}}, \bibinfo {author} {\bibfnamefont
  {G.}~\bibnamefont {Theureau}},  \emph {et~al.},\ }\href {\doibase 10/bck3wm}
  {\bibfield  {journal} {\bibinfo  {journal} {Mon. Not. Roy. Astron. Soc.}\
  }\textbf {\bibinfo {volume} {400}},\ \bibinfo {pages} {805} (\bibinfo {year}
  {2009})}\BibitemShut {NoStop}%
\bibitem [{\citenamefont {Fonseca}\ \emph {et~al.}(2014)\citenamefont
  {Fonseca}, \citenamefont {Stairs},\ and\ \citenamefont
  {Thorsett}}]{Fonseca:2014qla}%
  \BibitemOpen
  \bibfield  {author} {\bibinfo {author} {\bibfnamefont {E.}~\bibnamefont
  {Fonseca}}, \bibinfo {author} {\bibfnamefont {I.~H.}\ \bibnamefont {Stairs}},
  \ and\ \bibinfo {author} {\bibfnamefont {S.~E.}\ \bibnamefont {Thorsett}},\
  }\href {\doibase 10.1088/0004-637X/787/1/82} {\bibfield  {journal} {\bibinfo
  {journal} {Astrophys. J.}\ }\textbf {\bibinfo {volume} {787}},\ \bibinfo
  {pages} {82} (\bibinfo {year} {2014})}\BibitemShut {NoStop}%
\bibitem [{\citenamefont {Zhu}\ \emph {et~al.}(2019)\citenamefont {Zhu},
  \citenamefont {Desvignes}, \citenamefont {Wex}, \citenamefont {Caballero},
  \citenamefont {Champion}, \citenamefont {Demorest}, \citenamefont {Ellis},
  \citenamefont {Janssen}, \citenamefont {Kramer}, \citenamefont {Krieger}
  \emph {et~al.}}]{Zhu:2018etc}%
  \BibitemOpen
  \bibfield  {author} {\bibinfo {author} {\bibfnamefont {W.}~\bibnamefont
  {Zhu}}, \bibinfo {author} {\bibfnamefont {G.}~\bibnamefont {Desvignes}},
  \bibinfo {author} {\bibfnamefont {N.}~\bibnamefont {Wex}}, \bibinfo {author}
  {\bibfnamefont {R.}~\bibnamefont {Caballero}}, \bibinfo {author}
  {\bibfnamefont {D.}~\bibnamefont {Champion}}, \bibinfo {author}
  {\bibfnamefont {P.}~\bibnamefont {Demorest}}, \bibinfo {author}
  {\bibfnamefont {J.}~\bibnamefont {Ellis}}, \bibinfo {author} {\bibfnamefont
  {G.}~\bibnamefont {Janssen}}, \bibinfo {author} {\bibfnamefont
  {M.}~\bibnamefont {Kramer}}, \bibinfo {author} {\bibfnamefont
  {A.}~\bibnamefont {Krieger}},  \emph {et~al.},\ }\href {\doibase 10/ggth72}
  {\bibfield  {journal} {\bibinfo  {journal} {Mon. Not. Roy. Astron. Soc.}\
  }\textbf {\bibinfo {volume} {482}},\ \bibinfo {pages} {3249} (\bibinfo {year}
  {2019})}\BibitemShut {NoStop}%
\bibitem [{\citenamefont {Freire}\ \emph {et~al.}(2012)\citenamefont {Freire},
  \citenamefont {Wex}, \citenamefont {Esposito-Farese}, \citenamefont
  {Verbiest}, \citenamefont {Bailes}, \citenamefont {Jacoby}, \citenamefont
  {Kramer}, \citenamefont {Stairs}, \citenamefont {Antoniadis},\ and\
  \citenamefont {Janssen}}]{Freire:2012mg}%
  \BibitemOpen
  \bibfield  {author} {\bibinfo {author} {\bibfnamefont {P.~C.}\ \bibnamefont
  {Freire}}, \bibinfo {author} {\bibfnamefont {N.}~\bibnamefont {Wex}},
  \bibinfo {author} {\bibfnamefont {G.}~\bibnamefont {Esposito-Farese}},
  \bibinfo {author} {\bibfnamefont {J.~P.}\ \bibnamefont {Verbiest}}, \bibinfo
  {author} {\bibfnamefont {M.}~\bibnamefont {Bailes}}, \bibinfo {author}
  {\bibfnamefont {B.~A.}\ \bibnamefont {Jacoby}}, \bibinfo {author}
  {\bibfnamefont {M.}~\bibnamefont {Kramer}}, \bibinfo {author} {\bibfnamefont
  {I.~H.}\ \bibnamefont {Stairs}}, \bibinfo {author} {\bibfnamefont
  {J.}~\bibnamefont {Antoniadis}}, \ and\ \bibinfo {author} {\bibfnamefont
  {G.~H.}\ \bibnamefont {Janssen}},\ }\href {\doibase 10/f34tj8} {\bibfield
  {journal} {\bibinfo  {journal} {Mon. Not. Roy. Astron. Soc.}\ }\textbf
  {\bibinfo {volume} {423}},\ \bibinfo {pages} {3328} (\bibinfo {year}
  {2012})}\BibitemShut {NoStop}%
\bibitem [{\citenamefont {Ferdman}\ \emph {et~al.}(2014)\citenamefont
  {Ferdman}, \citenamefont {Stairs}, \citenamefont {Kramer}, \citenamefont
  {Janssen}, \citenamefont {Bassa}, \citenamefont {Stappers}, \citenamefont
  {Demorest}, \citenamefont {Cognard}, \citenamefont {Desvignes}, \citenamefont
  {Theureau} \emph {et~al.}}]{Ferdman:2014rna}%
  \BibitemOpen
  \bibfield  {author} {\bibinfo {author} {\bibfnamefont {R.~D.}\ \bibnamefont
  {Ferdman}}, \bibinfo {author} {\bibfnamefont {I.~H.}\ \bibnamefont {Stairs}},
  \bibinfo {author} {\bibfnamefont {M.}~\bibnamefont {Kramer}}, \bibinfo
  {author} {\bibfnamefont {G.~H.}\ \bibnamefont {Janssen}}, \bibinfo {author}
  {\bibfnamefont {C.~G.}\ \bibnamefont {Bassa}}, \bibinfo {author}
  {\bibfnamefont {B.~W.}\ \bibnamefont {Stappers}}, \bibinfo {author}
  {\bibfnamefont {P.~B.}\ \bibnamefont {Demorest}}, \bibinfo {author}
  {\bibfnamefont {I.}~\bibnamefont {Cognard}}, \bibinfo {author} {\bibfnamefont
  {G.}~\bibnamefont {Desvignes}}, \bibinfo {author} {\bibfnamefont
  {G.}~\bibnamefont {Theureau}},  \emph {et~al.},\ }\href {\doibase 10/f6mc33}
  {\bibfield  {journal} {\bibinfo  {journal} {Mon. Not. Roy. Astron. Soc.}\
  }\textbf {\bibinfo {volume} {443}},\ \bibinfo {pages} {2183} (\bibinfo {year}
  {2014})}\BibitemShut {NoStop}%
\bibitem [{\citenamefont {Weisberg}\ \emph {et~al.}(2008)\citenamefont
  {Weisberg}, \citenamefont {Stanimirovi{\'{c}}}, \citenamefont {Xilouris},
  \citenamefont {Hedden}, \citenamefont {de~la Fuente}, \citenamefont
  {Anderson},\ and\ \citenamefont {Jenet}}]{Weisberg2008}%
  \BibitemOpen
  \bibfield  {author} {\bibinfo {author} {\bibfnamefont {J.~M.}\ \bibnamefont
  {Weisberg}}, \bibinfo {author} {\bibfnamefont {S.}~\bibnamefont
  {Stanimirovi{\'{c}}}}, \bibinfo {author} {\bibfnamefont {K.}~\bibnamefont
  {Xilouris}}, \bibinfo {author} {\bibfnamefont {A.}~\bibnamefont {Hedden}},
  \bibinfo {author} {\bibfnamefont {A.}~\bibnamefont {de~la Fuente}}, \bibinfo
  {author} {\bibfnamefont {S.~B.}\ \bibnamefont {Anderson}}, \ and\ \bibinfo
  {author} {\bibfnamefont {F.~A.}\ \bibnamefont {Jenet}},\ }\href {\doibase
  10.1086/523345} {\bibfield  {journal} {\bibinfo  {journal} {Astrophys. J.}\
  }\textbf {\bibinfo {volume} {674}},\ \bibinfo {pages} {286} (\bibinfo {year}
  {2008})}\BibitemShut {NoStop}%
\bibitem [{\citenamefont {Cognard}\ \emph {et~al.}(2017)\citenamefont
  {Cognard}, \citenamefont {Freire}, \citenamefont {Guillemot}, \citenamefont
  {Theureau}, \citenamefont {Tauris}, \citenamefont {Wex}, \citenamefont
  {Graikou}, \citenamefont {Kramer}, \citenamefont {Stappers}, \citenamefont
  {Lyne} \emph {et~al.}}]{Cognard:2017xyr}%
  \BibitemOpen
  \bibfield  {author} {\bibinfo {author} {\bibfnamefont {I.}~\bibnamefont
  {Cognard}}, \bibinfo {author} {\bibfnamefont {P.~C.}\ \bibnamefont {Freire}},
  \bibinfo {author} {\bibfnamefont {L.}~\bibnamefont {Guillemot}}, \bibinfo
  {author} {\bibfnamefont {G.}~\bibnamefont {Theureau}}, \bibinfo {author}
  {\bibfnamefont {T.~M.}\ \bibnamefont {Tauris}}, \bibinfo {author}
  {\bibfnamefont {N.}~\bibnamefont {Wex}}, \bibinfo {author} {\bibfnamefont
  {E.}~\bibnamefont {Graikou}}, \bibinfo {author} {\bibfnamefont
  {M.}~\bibnamefont {Kramer}}, \bibinfo {author} {\bibfnamefont
  {B.}~\bibnamefont {Stappers}}, \bibinfo {author} {\bibfnamefont {A.~G.}\
  \bibnamefont {Lyne}},  \emph {et~al.},\ }\href
  {https://iopscience.iop.org/article/10.3847/1538-4357/aa7bee} {\bibfield
  {journal} {\bibinfo  {journal} {Astrophys. J.}\ }\textbf {\bibinfo {volume}
  {844}},\ \bibinfo {pages} {128} (\bibinfo {year} {2017})}\BibitemShut
  {NoStop}%
\bibitem [{\citenamefont {Deller}\ \emph {et~al.}(2018)\citenamefont {Deller},
  \citenamefont {Weisberg}, \citenamefont {Nice},\ and\ \citenamefont
  {Chatterjee}}]{Deller2018}%
  \BibitemOpen
  \bibfield  {author} {\bibinfo {author} {\bibfnamefont {A.~T.}\ \bibnamefont
  {Deller}}, \bibinfo {author} {\bibfnamefont {J.~M.}\ \bibnamefont
  {Weisberg}}, \bibinfo {author} {\bibfnamefont {D.~J.}\ \bibnamefont {Nice}},
  \ and\ \bibinfo {author} {\bibfnamefont {S.}~\bibnamefont {Chatterjee}},\
  }\href {\doibase 10.3847/1538-4357/aacf95} {\bibfield  {journal} {\bibinfo
  {journal} {Astrophys. J.}\ }\textbf {\bibinfo {volume} {862}},\ \bibinfo
  {pages} {139} (\bibinfo {year} {2018})}\BibitemShut {NoStop}%
\bibitem [{\citenamefont {Haniewicz}\ \emph {et~al.}(2020)\citenamefont
  {Haniewicz}, \citenamefont {Ferdman}, \citenamefont {Freire}, \citenamefont
  {Champion}, \citenamefont {Bunting}, \citenamefont {Lorimer},\ and\
  \citenamefont {McLaughlin}}]{Haniewicz:2020jro}%
  \BibitemOpen
  \bibfield  {author} {\bibinfo {author} {\bibfnamefont {H.~T.}\ \bibnamefont
  {Haniewicz}}, \bibinfo {author} {\bibfnamefont {R.~D.}\ \bibnamefont
  {Ferdman}}, \bibinfo {author} {\bibfnamefont {P.~C.}\ \bibnamefont {Freire}},
  \bibinfo {author} {\bibfnamefont {D.~J.}\ \bibnamefont {Champion}}, \bibinfo
  {author} {\bibfnamefont {K.~A.}\ \bibnamefont {Bunting}}, \bibinfo {author}
  {\bibfnamefont {D.~R.}\ \bibnamefont {Lorimer}}, \ and\ \bibinfo {author}
  {\bibfnamefont {M.~A.}\ \bibnamefont {McLaughlin}},\ }\href@noop {} {\
  (\bibinfo {year} {2020})},\ \Eprint {http://arxiv.org/abs/2007.07565}
  {arXiv:2007.07565 [astro-ph.SR]} \BibitemShut {NoStop}%
\end{thebibliography}%

\end{document}